\RequirePackage{silence}
\documentclass[useAMS,usenatbib]{mnras}
\usepackage[T1]{fontenc}
\usepackage{amsmath, amssymb, bm}
\usepackage[group-separator={,}]{siunitx}
\usepackage{physics}

\usepackage{graphicx}
\usepackage{caption}
\usepackage{subcaption}
\usepackage{stfloats}
\usepackage{booktabs}
\usepackage{tabularx}
\usepackage{multicol}

\usepackage[normalem]{ulem}
\usepackage{anyfontsize}

\usepackage{hyperref}
\hypersetup{breaklinks=true}
\usepackage[capitalize]{cleveref}
\usepackage{url}

\usepackage{xspace}
\usepackage{orcidlink}

\AtBeginDocument{\RenewCommandCopy\qty\SI}
\ExplSyntaxOn
\msg_redirect_name:nnn { siunitx } { physics-pkg } { none }
\ExplSyntaxOff

\DeclareRobustCommand{\appropto}{\mathrel{\vcenter{
    \offinterlineskip\halign{\hfil$##$\cr
        \propto\cr\noalign{\kern2pt}\sim\cr\noalign{\kern-2pt}}}}}

\newcommand{\kmsec}{\ensuremath{\mathrm{km}\,\mathrm{s}^{-1}}}
\newcommand{\kmsecMpc}{\ensuremath{\mathrm{km}\,\mathrm{s}^{-1}\,\mathrm{Mpc}^{-1}}}

\usepackage{fontawesome5}
\newcommand{\github}[1]{\href{#1}{\faGithubSquare}}
\newcommand{\githublink}{\github{https://github.com/harrydesmond/Volume_Prior/}}

\title[On the statistics of the distance ladder]{The subtle statistics of the distance ladder: On the distance prior and selection effects}
\author[H. Desmond et al]{
Harry Desmond$^{1}$\thanks{\href{mailto:harry.desmond@port.ac.uk}{harry.desmond@port.ac.uk}}\orcidlink{0000-0003-0685-9791}, Richard Stiskalek$^{2}$\orcidlink{0000-0002-0986-314X}, Jos\'e Antonio N\'ajera$^{1}$\orcidlink{0000-0001-9738-7704} and Indranil Banik$^{1}$\orcidlink{0000-0002-4123-7325}\\
$^{1}$Institute of Cosmology \& Gravitation, University of Portsmouth, Dennis Sciama Building, Portsmouth, PO1 3FX, UK\\
$^{2}$Astrophysics, University of Oxford, Denys Wilkinson Building, Keble Road, Oxford, OX1 3RH, UK}

\date{Accepted XXX. Received YYY; in original form ZZZ}
\pubyear{\the\year{}}

\begin{document}
\label{firstpage}
\pagerange{\pageref{firstpage}--\pageref{lastpage}}
\maketitle

\begin{abstract}
	Statistical methodology is rarely considered significant in distance-ladder studies or a potential contributor to the Hubble tension. We suggest it should be, highlighting two appreciable issues. First, astronomical distances are inferred latent parameters, requiring a prior. We show that the (often implicit) uniform priors on distance moduli common to Bayesian distance-ladder analyses bias distances low due to objects being uniformly distributed in volume, which biases the Hubble constant high. Frequentist $\chi^2$ methods are unbiased for volume- or redshift-limited samples only if the redshift uncertainty (including peculiar velocities) vanishes, though simulation-based calibration can correct the bias. Second, in a Bayesian framework, selection effects introduce additional posterior factors describing the probability of objects entering the sample under the model. These partly counteract the volume prior, depending on the nature of the selection. After detailed analytic and mock-based studies, we quantify the volume-prior effect in the CosmicFlows-4 and SH0ES samples. Both use frequentist methods, so the effect appears as a potential estimator bias rather than a missing prior. The implied Hubble constant shifts are significant but must not be applied na\"{\i}vely---principled selection modelling is also required, as we investigate explicitly for CosmicFlows-4. Both effects should already be captured by the SH0ES pipeline's simulation-based bias corrections. Our work highlights the crucial need to model both distances and selection accurately, either directly in a Bayesian forward model, or via post-hoc simulation-based corrections with realistic source and selection distributions. Such modelling requires samples with known, homogeneous selection criteria, which future surveys should prioritise.
\end{abstract}

\begin{keywords}
    cosmology: distance scale -- galaxies: distances and redshifts -- cosmological parameters -- methods: statistical -- methods: numerical
\end{keywords}

\section{Introduction}

A key goal of modern cosmology is to infer accurate distances. The distance--redshift relation (or Hubble diagram) describes the present-day expansion rate $H_0$ and the low-$z$ deceleration parameter $q_0$. Measuring the Hubble diagram precisely is particularly pressing in light of the ``Hubble tension'' \citep{Valentino_2025}, a $\ga 5\sigma$ mismatch between $H_0$ inferred from the local distance ladder---specifically the \emph{Supernovae and H0 for the Equation of State} pipeline (SH0ES;~\citealt{Riess_2022,Breuval_2024})---versus that reconstructed assuming $\Lambda$CDM from the Cosmic Microwave Background (CMB) anisotropies as measured by the \textit{Planck} satellite~\citep{Planck_2020_cosmo, Tristram_2024} or from the ground \citep{Calabrese_2025, Camphuis_2025}. The distance--redshift relation can also be used to distinguish between competing gravitational or cosmological models~\citep{H0_F5, Anton_Clifton, Stiskalek_2025_void}, measure velocity flows and hence the cosmography of the local Universe~\citep{Dupuy_Courtois}, infer the local growth rate of structure~\citep{VFO}, and test the Cosmological Principle~\citep{Watkins,Stiskalek_anisotropy,Yasin}.

This paper focuses on two distinct but interrelated intricacies in the statistics of the distance ladder: the priors imposed on galaxy distances and the modelling of selection effects. We will show these to have a potentially significant impact on inferred quantities like $H_0$, such that inappropriate modelling can lead to significant biases. These issues must therefore be carefully addressed in distance-ladder pipelines for precise constraints to be accurate.

Distances cannot be measured directly. They must be inferred as (latent) model parameters given observables pertaining to them. These typically relate to ``standard candles'' or ``standard rulers'', astrophysical objects whose absolute brightness or length scale can be calibrated. In conjunction with a measured relative brightness (flux) or scale (angular size), this enables inference of distances to the objects.
Distances therefore require priors: the probability distributions we expect for them before any observation is carried out. It is sometimes considered that ``uninformative'' priors, which allow the posterior to be determined by the data likelihood, are uniform or flat. In some cases, a flat prior is reasonable for want of a better assumption, but in others, the physics of the situation (reflected in the data-generation process) dictates the ``correct'' prior, namely the one that leads to an unbiased inference. The distance $r$ is such a case: since objects are intrinsically uniformly distributed in space and space is three-dimensional, the distribution of objects' distances increases as $r^2$. The first main aim of this paper is to show that this requires one's prior on distance to also go as $r^2$, and to quantify the bias induced by any other choice.

This is not new. The first exposition of the uniform-in-volume prior (henceforth ``volume prior'') effect dates back to~\citet{Eddington1914}; this was then quantified and formalised by~\citet{Malmquist_1922}, for which reason it is often referred to as ``homogeneous Malmquist bias''. This is something of a misnomer because it is not a bias if one accounts for it correctly, but in this sense anything is a bias. Further confusion arises because Malmquist bias is often considered as a \emph{selection effect} which only kicks in when one has e.g. a flux-limited survey, leading to preferential detection of intrinsically brighter objects. This obscures the fact that \emph{even without any selection at all} (i.e. for a volume-limited survey), objects are likely to be more distant than where the observational likelihood peaks. If this peak is considered the `measured' value without accounting for the volume prior, one will infer distances that are biased low. Given the measured redshifts, the Hubble constant will then be biased high. This is because it is more likely that the observed magnitude (using a standard candle as an example) scattered down from the true apparent magnitude than vice versa, because there are more galaxies at larger distance where apparent magnitudes are higher \citep[careful descriptions can be found in][]{Lynden_Bell_1988, Strauss_Willick, Lavaux_2015}.

Unfortunately, some state-of-the-art distance ladder analyses neglect the volume prior. This appears to stem from a combination of (1) treating distances as observables rather than model parameters, leading to the erroneous assumption that they do not require priors, and (2) working in a frequentist context, leading to the erroneous assumption that one does not need priors at all. There may also be a practical component: if latent distances cannot be integrated out analytically, they must either be marginalised over numerically or sampled, but most Bayesian inference algorithms cannot handle hundreds or thousands of parameters. The result is that inferred distances reflect only the data likelihood, implicitly assuming a flat prior on distance ($r^0$), or, more commonly, on the distance modulus $\mu$. Since $\mu$ is linearly related to $\log(r)$, this corresponds to a $1/r$ prior, exacerbating the bias. In the limit of negligible redshift uncertainty (including no peculiar velocities), a linear distance--redshift relation and no inhomogeneous Malmquist modelling, frequentist $\chi^2$ optimisation is equivalent to Bayesian forward-modelling under a flat-$\mu$ prior. However, we will see that this equivalence comes apart when modelling selection effects within the Bayesian methodology.

While ``homogeneous Malmquist bias'' is not a selection effect, it does have an interesting interplay with selection effects. Such effects arise if (as is typically the case) one does not include in one's sample all objects within some known physical volume, but rather (implicitly or explicitly) includes or excludes objects on the basis of an observable quantity such as apparent magnitude or redshift. This preferentially selects for nearer objects, which counteracts to some extent the preference of the volume prior for objects to be more distant. Selection effects are accounted for in various ways in the literature, the most common method being through the use of simulations. Here we adopt a Bayesian forward-modelling approach, deriving the additional factors that must be included in the posterior for various types of selection. In particular, we find that assuming a redshift limit, a fortuitous cancellation means that a uniform-in-$\mu$ distance prior without accounting for selection produces an unbiased estimate of $H_0$ in the absence of redshift uncertainties, higher-order cosmographic terms, and anisotropies arising from e.g. peculiar velocities. In general, the method is significantly biased. Further, selection effects can only be accounted for in an a-priori principled way through Bayesian forward-modelling, which constructs an inference that matches the causal mechanism generating the data.

Our results imply that a level of bias is present in a great many distance-ladder studies, including both frequentist $\chi^2$ \citep[e.g.][]{CF4, Dhawan, Burns, Schombert, Blakeslee, Jaegar, Freedman_2025} and Bayesian forward-modelling \citep[e.g.][]{March, Feeney, Becker, Suvodip, BSN_1} analyses. Selection effects are scantily treated in all of these studies, with the Bayesian analyses all failing to implement a volume prior. Whether the bias is towards a lower or higher $H_0$ depends on the exact fitting method used and selection effects in the sample. We will calculate the volume prior effect quantitatively in CosmicFlows-4 \citep[CF4;][]{CF4} and SH0ES~\citep{Riess_2022}, for the former also quantifying the impact of selection.

The structure of the paper is as follows. In Sec.~\ref{sec:dist_prior}, we quantify the effect of the distance prior in a simplified distance-ladder setup for a volume-limited sample; first analytically in the case of negligible redshift uncertainty, and then numerically in general.
Sec.~\ref{sec:selection} studies magnitude and redshift selection effects, calculating bias in various methodologies in each case.
In Sec.~\ref{sec:cf4}, we apply this to the real-world CF4 sample, demonstrating the significant biases that can be produced by model mis-specification. Sec.~\ref{sec:sh0es} is devoted to the SH0ES data, where we show that the volume prior is an $\approx1.7\sigma$ effect on $H_0$ that should already be accounted for in the simulation-based SH0ES pipeline. Further discussion and generalisation may be found in Sec.~\ref{sec:conc}. Appendix~\ref{sec:appendix} provides further analytic detail in the case of negligible redshift uncertainty and explores the opposite limit of negligible magnitude uncertainty.

\section{The distance prior effect for a volume-limited sample}\label{sec:dist_prior}

We consider a simple toy setup for inference of $H_0$ through the distance ladder. Initially we restrict attention to Bayesian inference where distances are treated as latent parameters; in Sec.~\ref{sec:chi2_method} we instead consider frequentist $\chi^2$ inference in which distances are mapped deterministically to redshifts through Hubble's law.

\begin{figure}
	\centering
	\includegraphics[width=\columnwidth]{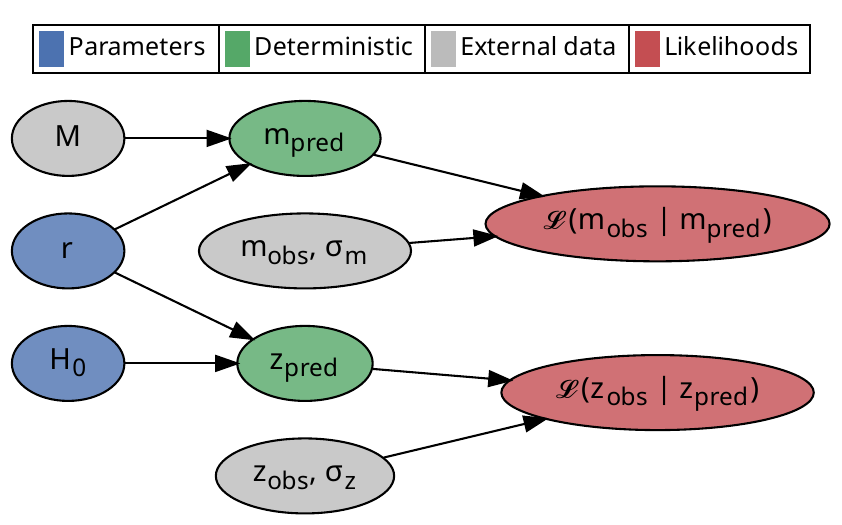}
	\caption{Directed acyclic graph depicting the simple distance ladder inference we use to illustrate the effect of the distance prior.}
	\label{fig:DAG}
\end{figure}

Suppose $N$ observed galaxies are known to have latent distances between $r_\text{min}$ and $r_\text{max}$, corresponding to a \emph{volume-limited sample}. Although the distances are not known, we suppose $r_\text{min} > 0$ and $r_\text{max}$ are known. We measure the redshifts $z_i$ with constant Gaussian redshift uncertainties $\sigma_z$, which would in practice arise mainly from peculiar velocities. Each galaxy contains a standard candle of known absolute magnitude $M$ (for illustration---any other type of distance indicator would behave the same). We also measure the apparent magnitudes $m_i$ with constant Gaussian uncertainty $\sigma_m$. The unknown parameters are the distance $r_i$ to each galaxy $i$ and the Hubble parameter $H_0$.\footnote{We assume the galaxies are sufficiently close for higher-order terms in the cosmographic expansion to be irrelevant, comoving and luminosity distances to be equal, and geometric effects from non-Euclidean space to be unimportant. These complicate the calculation but do not change the story.} This enables us to predict the observables:
\begin{align}\label{eq:observables}
    z_{\mathrm{pred},i} &= \frac{H_0 r_i}{c}, \\
    m_{\mathrm{pred},i} &= M + \mu_i,
\end{align}
where the distance modulus to galaxy $i$ is defined as
\begin{equation}
    \mu_i \equiv \alpha \ln(r_i), \, \quad \alpha \equiv 5/\ln 10.
\end{equation}
For simplicity this neglects the conventional and unimportant ``+25'' in the definition of $\mu_i$ (i.e., to use our convention, 25 should be subtracted from distance moduli as normally defined). Alternatively, the distances can be considered to be in units of 10~pc. The directed acyclic graph of this setup is shown in Fig.~\ref{fig:DAG}.

We assume a wide uniform prior on $H_0$ and a prior $\pi(r)$ on all the $r_i$. We parametrise $\pi(r)$ as a power-law with exponent $k$, i.e.
\begin{equation}\label{eq:prior}
    \pi(r_i) =
    \begin{cases}
        C \: r_i^k, & r_{\min} \leq r_i \leq r_{\max}, \\
        0, & \text{otherwise},
    \end{cases}
\end{equation}
where $C$ is the normalisation constant. This is given by
\begin{equation}\label{eq:prior_C}
    C =
    \begin{cases}
    \dfrac{k+1}{\,r_{\max}^{k+1} - r_{\min}^{k+1}\,}, & k \neq -1, \\[1.2em]
    \dfrac{1}{\ln \left(r_{\max} / r_{\min}\right)}, & k = -1.
\end{cases}
\end{equation}
This reflects the defining feature of a volume-limited sample that all true distances lie between the known limits $r_\text{min}$ and $r_\text{max}$.\footnote{For simplicity we neglect the inhomogeneous Malmquist contribution; this would modify the prior to $\pi(r_i) \propto n(r_i,\,\bm{u}_i)\,r_i^k$, where $n(r_i,\,\bm{u}_i)$ is the real-space number density of sources at distance $r_i$ along line of sight $\bm{u}_i$. This could be modelled using a reconstructed density field such as that of~\citet{Carrick_2015} or~\citet{McAlpine_2025}.}

\subsection{Analytic sketch}\label{sec:analytic}

Up to an additive constant, the negative log-posterior is
\begin{equation}\label{eq:nlp_full}
    \begin{split}
        -\ln \mathcal{P}(\{r_i\},\,H_0\mid& \{z_i,\,m_i\}) =
        \sum_{i=1}^N \left\{\frac{(M + \alpha \ln r_i - m_i)^2}{2 \sigma_m^2} \right.\\
        &+ \left.\frac{(H_0 r_i / c - z_i)^2}{2 \sigma_z^2}
        - k \ln r_i\right\}
    \end{split}
\end{equation}
if $\{r_i\} \in [r_{\rm min},\,r_{\max}]$ and infinite otherwise.

We suppose here that the redshift uncertainties are small. The specific condition compares the redshift-derived distance uncertainty $c\sigma_z/H_0$ to the magnitude-derived one $r\sigma_m/\alpha$, so that we are assuming
\begin{equation}\label{eq:small_sigmaz}
	\frac{c\sigma_z}{H_0} \ll \frac{r\sigma_m}{\alpha}
\end{equation}
(see also Appendix~\ref{sec:appendix}).
In this case, distances are deterministically related to the redshifts given an assumed $H_0$, replacing the second term in the log-posterior with the $\delta$-function constraint $r_i = c z_i/H_0$. Switching variable from $r$ to $\mu$, we can write
\begin{equation}\label{eq:simple_nlp}
    -\ln\mathcal{P}(\{\mu_i\},\,H_0\mid \{m_i\}) = \sum_{i=1}^N \left\{\frac{(\mu_i - m_i + M)^2}{2 \sigma_m^2} - \frac{k+1}{\alpha} \mu_i \right\},
\end{equation}
where the ``+1'' in the final term derives from the Jacobian of the $r\rightarrow\mu$ parametrisation, and again provided that all $r_i$ are between $r_\text{min}$ and $r_\text{max}$. Note that $\{\mu_i\}$ and $H_0$ are not independent variables given the $\delta$-function constraint: there is only one independent degree of freedom because for given $H_0$, all the $\mu_i$ follow from the measured redshifts. Thus one can vary with respect to \emph{either} $\{\mu_i\}$ \emph{or} $H_0$, but not both. The way in which we have written the posterior in Eq.~\ref{eq:simple_nlp} invites variation with respect to $\{\mu_i\}$.

Differentiating with respect to $\mu_i$, we find that the posterior peaks at
\begin{eqnarray}\label{eq:simple_muhat}
    \hat{\mu}_i ~=~ m_i - M + \frac{(k+1)}{\alpha} \sigma_m^2,
    \label{mu_VP_corrected}
\end{eqnarray}
where we use a hat to denote a maximum a posteriori (MAP) value. The difference in $\ln \hat{r}$ using different choices of $k$ is
\begin{eqnarray}\label{eq:simple_deltamuhat}
    \Delta \ln \hat{r} ~=~ \frac{\sigma_m^2}{\alpha^2} \Delta k.
    \label{Ln_r_bias}
\end{eqnarray}
As the inferred $H_0$ is inversely proportional to the distances from the $\delta$-function constraint, this implies an opposite shift in $\hat{H}_0$ by
\begin{eqnarray}\label{eq:simple_deltaH0}
    \Delta \ln \hat{H}_0 ~=~ -\frac{\sigma_m^2}{\alpha^2} \Delta k.
    \label{Delta_Ln_H0}
\end{eqnarray}
Thus, a survey analysing a population of objects with known cosmological redshifts and absolute magnitude but $\sigma_m = 0.1$~mag would have a 0.64\% shift in $\hat{H}_0$ for $\Delta k=3$, corresponding to the difference between the uniform-in-$\mu$ prior ($k=-1$) and the uniform-in-volume prior ($k=2$). This is independent of the sample size. Since $\ln \mathcal{P}$ in Eq.~\ref{eq:simple_nlp} remains Gaussian in $\mu$ even after the volume prior is considered, it does not alter the uncertainty in the posterior inference on $\mu$. This means that one can account for the volume prior in algorithms designed to use a uniform prior in $\mu$ simply by increasing the distance moduli by the amount given in Eq.~\ref{mu_VP_corrected}.

Our results can readily be generalised to the case where the $\sigma_{m,i}$ are not all the same. Since the relative statistical weight of any observation $\propto \sigma_{m,i}^{-2}$ and this factor precisely cancels the $\sigma_{m,i}^2$ factor in the bias, Eq.~\ref{Delta_Ln_H0} would become
\begin{eqnarray}\label{eq:simple_deltaH0_varerr}
    \Delta \ln \hat{H}_0 ~=~ -\frac{\Delta k}{\alpha^2 \langle \sigma_m^{-2} \rangle} \, .
    \label{Bias_H0_unequal_mu}
\end{eqnarray}

It is also instructive to calculate the bias relative to the $H_0$ posterior width. Since we can infer $H_0$ with the distance to any single object, we expect that
\begin{eqnarray}\label{eq:simple_H0std}
    \sigma \left( \ln H_0 \right) ~=~ \frac{\sigma \left( \ln r \right)}{\sqrt{N}} ~=~ \frac{\sigma_m}{\alpha\sqrt{N}} \, ,
\end{eqnarray}
where $\sigma \left( X \right)$ is the posterior uncertainty on any quantity $X$. We can then divide the expected bias in $\ln H_0$ (Eq.~\ref{eq:simple_deltaH0}) by our estimated $\sigma \left( \ln H_0 \right)$ to get the expected relative bias:
\begin{eqnarray}\label{eq:simple_deltaH0rel}
\frac{\Delta \ln \hat{H}_0}{\sigma(\ln H_0)} = \frac{\Delta \hat{H}_0}{\sigma(H_0)} =  -\frac{\sigma_m}{\alpha} \sqrt{N} \Delta k,
\label{Bias_analytic}
\end{eqnarray}
which generalises to $\sqrt{N} \: \Delta k \: \langle \sigma_m^{-2} \rangle^{-1/2} / \alpha$ for variable $\sigma_m$. If $\sigma_m=0.1$~mag, $N=2000$, and $\Delta k=3$, the bias is 6.2$\sigma$.

For an alternative, more rigorous derivation of these results which also investigates the opposite limit of large $\sigma_z$, see Appendix~\ref{sec:appendix}.

\subsection{Mock data tests}\label{sec:mock}

Sec.~\ref{sec:analytic} shows that the distance prior makes a difference for $H_0$, but leaves two questions unanswered. The first, more important one is which prior actually leads to an unbiased inference of $H_0$, i.e. produces a posterior centred around the true value? The second is the extent to which the results depend on the approximations employed for the calculation to be analytically tractable. In Sec.~\ref{sec:analytic} this is the assumption that $\sigma_z\rightarrow 0$, while in Appendix~\ref{sec:appendix} we also consider the opposite limit in which $\sigma_m \rightarrow 0$. We address both of these issues in this section by generating mock data, inferring $H_0$ and the $r_i$ using Markov Chain Monte Carlo, and investigating biases in the posterior relative to the known truths. Since we are now treating general $\sigma_z$, we use the posterior of Eq.~\ref{eq:nlp_full}.

The mocks are generated according to
\begin{equation}\label{eq:mockgeneration}
\begin{split}
\bar{r}& \hookleftarrow \left[r_\text{min}^3 + (r_\text{max}^3 - r_\text{min}^3) \: \mathcal{U}(0,1)\right]^\frac{1}{3},\\
m      &\hookleftarrow \mathcal{N}\!\left(M + \mu(\bar{r}), \sigma_m\right),\\
z      &\hookleftarrow \mathcal{N}\!\left(\bar{H}_0 \bar{r}/c, \sigma_z\right),
\end{split}
\end{equation}
where an overbar denotes the true (generating) value of a parameter, $\mathcal{U}(a,b)$ denotes a uniform distribution between $a$ and $b$, and $\mathcal{N}(x,\sigma)$ denotes a normal distribution of mean $x$ and standard deviation $\sigma$. For illustration we take $r_\text{min} = 5$ Mpc, $r_\text{max} = 100$ Mpc, $\sigma_m=0.1$, $\sigma_z=0.001$ (corresponding to a peculiar velocity uncertainty of $300~\kmsec$), $M=-5$, $\bar{H}_0 = 70~\kmsecMpc$, and $N_\text{gal}=2000$ objects. We then infer $H_0$ with a wide uniform prior and $r$ with the prior given by Eqs.~\ref{eq:prior} and~\ref{eq:prior_C}. As there are 2001 parameters, we employ the No U-Turns Sampler method of Hamiltonian Monte Carlo, as implemented in \texttt{NumPyro}~\citep{Hoffman_2011, Phan_2019, Bingham_2019}, with sufficient steps to produce a Gelman--Rubin statistic~\citep{Gelman_1992} $<1.01$ in all cases. Note that for this setup, the left hand side of Eq.~\ref{eq:small_sigmaz} is 4.29 for $H_0=70~\kmsecMpc$ (at $\langle r \rangle = 75$ Mpc), while the right hand side is 3.45. Redshift and magnitude uncertainties therefore have comparable effects, so we cannot expect either limiting case to be accurate.

\begin{figure}
    \centering
    \includegraphics[width=\columnwidth]{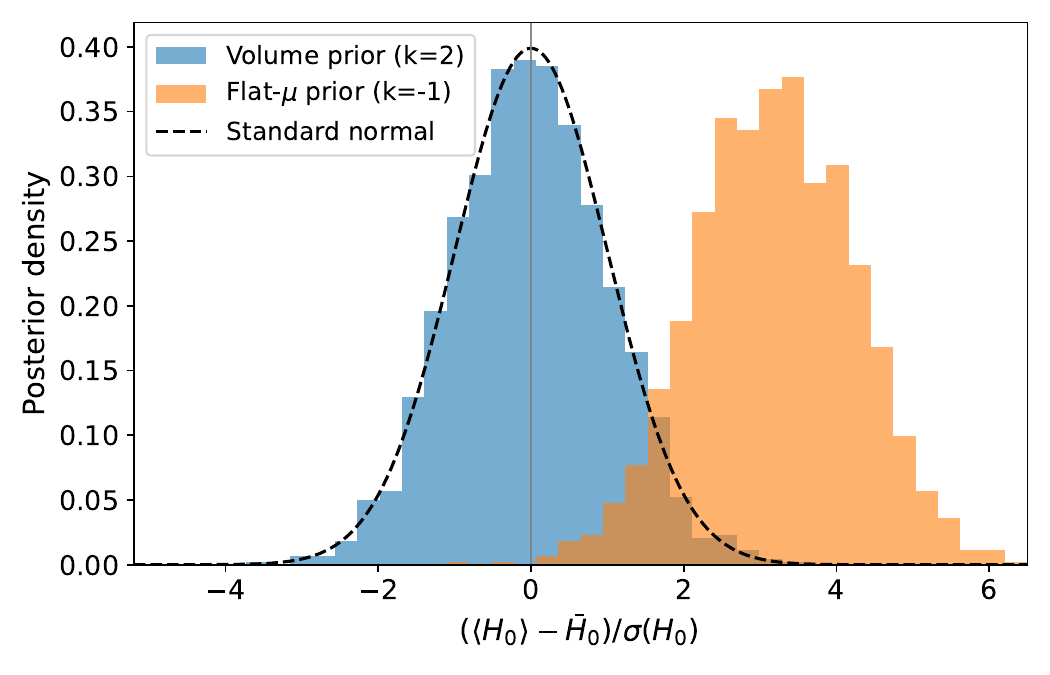}
    \caption{The relative bias in $H_0$ (in units of $\sigma$; Eq.~\ref{eq:mockbiasdef}) produced by the uniform-in-volume and uniform-in-$\mu$ distance priors across 1500 mock datasets. The agreement of the volume prior with the standard normal distribution shows that it is unbiased.}
    \label{fig:bias_var}
\end{figure}

We generate and fit 1500 mock datasets, varying only the random numbers used in Eq.~\ref{eq:mockgeneration}. For each dataset, we summarise the nearly Gaussian posterior on $H_0$ by its mean $\langle H_0 \rangle$ and standard deviation $\sigma(H_0)$ across the Monte Carlo samples. We then quantify the relative bias by
\begin{equation}\label{eq:mockbiasdef}
    \mathcal{B}(H_0) \equiv \frac{\langle H_0 \rangle - \bar{H}_0}{\sigma(H_0)}.
\end{equation}
This should have a standard normal distribution across the mock datasets if the model is unbiased, i.e. the $X$ per cent credible interval contains the true value $X$ per cent of the time for all $X$. Fig.~\ref{fig:bias_var} shows the distribution of $\mathcal{B}(H_0)$ for both $\pi(r) \propto 1/r$ ($k=-1$) and $\pi(r) \propto r^2$ ($k=2$). It is clear that the former is biased and the latter unbiased, reflecting the fact that the tracers are uniformly distributed in volume. The uniform-in-$\mu$ prior model is mis-specified. (Recall that we are still working entirely within a Bayesian context; frequentist $\chi^2$ methods will be discussed below.)

We see that in this setup, the uniform-in-$\mu$ prior produces an $\approx 3.2\sigma$ bias on average, which is smaller than the $6.2\sigma$ derived above. This is partly because the assumption of negligible $\sigma_z$ breaks down, causing Eq.~\ref{eq:simple_deltaH0} to be inaccurate (it predicts $\Delta \hat{H}_0 = 0.45$ km/s/Mpc, but the average across the mock datasets is 0.35). Moreover, $\sigma(H_0)$ is increased by the redshift uncertainties, lowering the relative bias. We note that the prior exponent $k$ can be inferred as a hyperparameter of the model, yielding $\approx2\pm0.07$ on each dataset in this setup.

\begin{figure}
	\centering
	\includegraphics[width=\columnwidth]{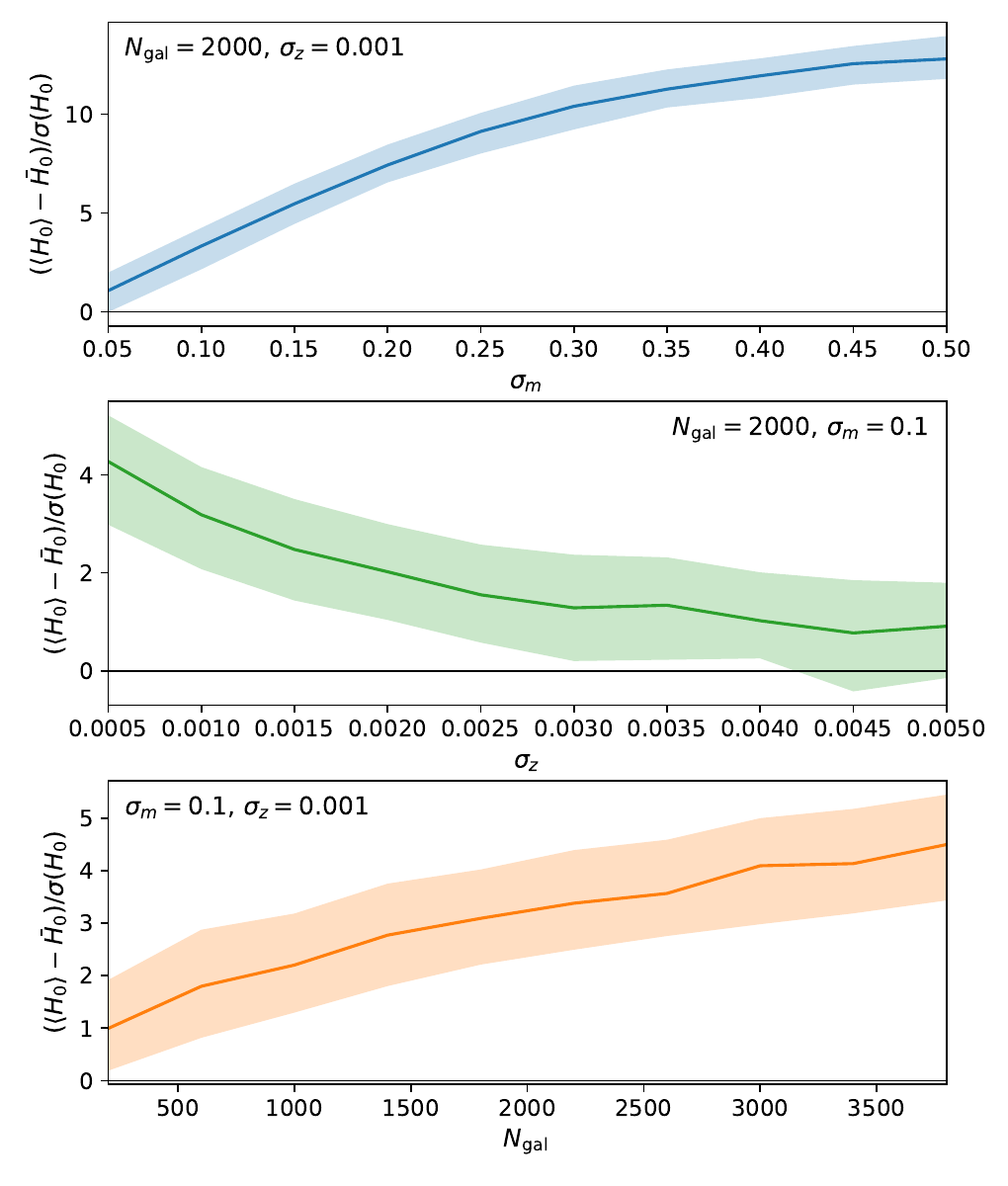}
	\caption{The average relative bias (solid lines) and 16$^\text{th}$-84$^\text{th}$ percentile range (shaded bands) produced by the uniform-in-$\mu$ prior for a volume-limited sample as a function of $\sigma_m$, $\sigma_z$, and $N_\text{gal}$ for fixed fiducial values of the other parameters, as indicated on each panel.}
	\label{fig:bias}
\end{figure}

To illustrate the dependence of the bias on the number of galaxies in the sample and the sizes of the uncertainties, we show in Fig.~\ref{fig:bias} the bias across many mock datasets as a function of these parameters separately, fixing the other two parameters to $\sigma_m=0.1$, $\sigma_z=0.001$, and $N_\text{gal}=2000$. We see that the main driver of a high relative bias is the high magnitude uncertainty (or more generally uncertainty from the distance indicator), which is exacerbated by a small redshift uncertainty and a large dataset.

\subsection{The $\chi^2$ estimator}\label{sec:chi2_method}

In frequentist inference, distances are not treated as latent parameters but rather mapped deterministically to redshifts through the assumed cosmography. This utilises Eq.~\ref{eq:simple_nlp} without the second term on the right hand side (or equivalently Eq.~\ref{eq:post_H0_k} without the denominator), thereby assuming $k=-1$.
There is no mechanism to account for selection effects within this framework because a generative model for the data is not constructed.
For a volume-limited sample, accounting for selection effects means requiring distances to lie within the assumed-known distance bounds that define the sample. Since the $\chi^2$ method does not do this, its results will generally differ from the Bayesian flat-$\mu$ case even under a pure volume limit. This applies even for $\sigma_z\rightarrow0$ where the distance bounds can be violated for some $H_0$ ranges: these are therefore excluded by the Bayesian method but not by $\chi^2$.

The deterministic mapping between distance and redshift breaks down for $\sigma_z > 0$, requiring the $\chi^2$ estimator to be modified. Without doing so, its bias can be shown to be
\begin{equation}\label{eq:chi2_bias_volume}
    \langle H_0 \rangle - \bar{H}_0
    = -\frac{\left(c \sigma_z\right)^2}{2 \bar{H}_0}\,
    \langle r^{-2} \rangle,
\end{equation}
for a volume-limited sample (and a uniform-in-volume population distribution). Typically, the finite $\sigma_z$ is propagated linearly to the magnitude likelihood to obtain an effective total uncertainty
\begin{equation}\label{eq:chi2_sigmaz_propagation}
    \sigma_\text{tot}^2 = \sigma_m^2 + \left(\frac{ \alpha\sigma_z}{z}\right)^2
\end{equation}
which replaces $\sigma_m$ in the denominator of the $\chi^2$. The estimator approaches the deterministic limit only when the second variance term is subdominant, which is the condition of Eq.~\ref{eq:small_sigmaz}. Eq.~\ref{eq:chi2_sigmaz_propagation} introduces another difference from the Bayesian methods, which replace linear error propagation with marginalisation over latent distances in a forward model for the observables. We will investigate the bias of this under various selection assumptions in Sec.~\ref{sec:full_bias_results}. As these biases are intrinsic to the estimator, they cannot be corrected by making the prior imply a more accurate generative model for the data---after all, there is no prior. The biases must therefore be ameliorated with post-hoc bias correction schemes.

\section{The impact of selection}\label{sec:selection}

So far we have neglected the issue of selection effects in the distance-ladder data by assuming a volume-limited sample. In practice, however, selection effects play an important role and must be modelled to achieve an unbiased inference. In this section we derive the effect of redshift, magnitude, and redshift-plus-magnitude selection in the Bayesian forward modelling context. We also present a phenomenological selection model that modifies the distance prior, and relate our results to the $\chi^2$ estimator often used for inferring $H_0$.

To understand the interplay between the distance prior and selection effects, it is helpful to make explicit the factors of $H_0$ introduced by a volume prior. We noted below Eq.~\ref{eq:simple_nlp} that the posterior could be written in terms of either $\{\mu_i\}$ or $H_0$, which are not independent given the $\delta$-function constraint (Eq.~\ref{eq:delta}). In terms of $H_0$, Eq.~\ref{eq:simple_nlp} reads
\begin{equation}
    -\ln\mathcal{P} = \sum_i \left\{ \frac{(\alpha\ln(c z_i/H_0) - m_i + M)^2}{2\sigma_m^2} - (k+1) \ln\left(\frac{c z_i}{H_0}\right) \right\},
\end{equation}
which implies
\begin{equation}\label{eq:post_H0_k}
    \mathcal{P}(H_0 \mid \{z_i,\,m_i\}) \propto \prod_i \frac{\mathcal{N}(m_i | \alpha \ln \hat{r}_i + M,\,\sigma_m^2)}{H_0^{1 + k}}.
\end{equation}
The choice of $k = -1$ corresponds to the ``$\chi^2$'' case discussed further below, which simply maximises the likelihood of the observed magnitudes. Note that this equation assumes \emph{no} bounds on the $\hat{r}_i$, violating the volume-limited assumption that the $r_\text{min}$ and $r_\text{max}$ of the sample are known \textit{a priori}. This helps to make contact with cases that do have a non-trivial selection, where these $r$ bounds are replaced by additional factors in the posterior.

Bayesian modelling of selection effects was pioneered by~\citet{Kelly_2007,Kelly_2008}; while this was taken up in the gravitational wave community (e.g.~\citealt{Mandel,Mortlock,Gair,Palmese}) it obtained little traction in other fields (although see~\citealt{Buchner}). Recently it has been adapted for distance ladder modelling in~\citet{CH0, Stiskalek_MW}. The key point is that in the presence of selection, the posterior on population parameters $\bm{\Lambda}$ (not object-specific latent parameters) given data $\bm{d}_\text{obs}$ is modified to
\begin{equation}\label{eq:posterior_with_selection}
    \mathcal{P}(\bm{\Lambda}\mid \bm{d}_{\rm obs}) \propto \pi(\bm{\Lambda}) [p(S=1\mid\bm{\Lambda})]^{-N} \: \prod_{i=1}^{N} \mathcal{L}(\bm{d}_i\mid\bm{\Lambda}).
\end{equation}
The first and third factors on the right hand side describe the regular posterior in the absence of selection. $p(S=1 \mid \bm\Lambda)$ is the probability for a randomly drawn object from the population described by $\bm\Lambda$ to pass the selection cut, while as before $N$ is the number of objects in the sample (note that this was denoted $n$ in~\citealt{CH0}, where $N$ instead denoted the total number of sources, observed or not). This is derived by marginalising over the unobserved data in the full population~\citep{Kelly_2008}. Here $\bm{d}_\text{obs} = \{m_i, z_i\}$ and $\bm\Lambda = H_0$.

\subsection{Redshift selection}\label{sec:sel_z}

In the case of redshift selection, we have
\begin{equation}
    p(S=1\mid H_0)
    =
    \iint \dd z\,\dd r\,p(S = 1 \mid z) \mathcal{L}(z\mid r,\,H_0)\,\pi(r),
\end{equation}
where $p(S = 1 \mid z)$ is a selection indicator given some source redshift $z$, while $\mathcal{L}(z\mid r,\,H_0)$ is a likelihood of the redshift given some source distance $r$ and $H_0$. To make the dependence on $H_0$ explicit, we isolate its factors within the integral. Introducing a change of variable $x = H_0 r$, the expression becomes
\begin{equation}\label{eq:zsel}
\begin{split}
    p(S{=}1\mid H_0)
    &\propto
    \frac{1}{H_0^{1+k}}
    \iint \dd z\,\dd x\;
    p(S{=}1\mid z)\,
    \exp\!\left[-\frac{(z-x)^2}{2\sigma_z^2}\right]\,
    x^{k}\\
    &\propto \frac{1}{H_0^{1+k}},
\end{split}
\end{equation}
where the factors of $H_0^{-1}$ and $H_0^{-k}$ arise from $\dd x = H_0\,\dd r$ and the $r^k$ distance prior, respectively. The second proportionality follows because the integral is independent of $H_0$.
This assumes that $\sigma_z$ is constant, but holds even if $\sigma_z \rightarrow 0$.

If we now explicitly assume that $\sigma_z \rightarrow 0$, combining the redshift-selection results with Eqs.~\ref{eq:post_H0_k} and~\ref{eq:posterior_with_selection} implies that
\begin{equation}\label{eq:zsel_chi2}
    \mathcal{P}(H_0 \mid \{z_i,\,m_i\})
    \propto
    \prod_i
    \mathcal{N}(m_i | \alpha \ln \hat{r}_i + M,\,\sigma_m^2).
\end{equation}
(If $\sigma_z > 0$, the factors of $H_0$ can no longer be factored out as in Eq.~\ref{eq:post_H0_k}, so they do not cancel with the selection term and the posterior becomes dependent on the assumed $k$.)
This remarkable result indicates that, under the given assumptions, the modification to the posterior induced by redshift selection exactly cancels the distance prior under the assumption of a power-law distance prior. Since Eq.~\ref{eq:zsel} shows that this modification vanishes for $k=-1$, one can therefore get away with not accounting for selection if one uses an unphysical uniform-in-$\mu$ prior, a rare instance of two mistakes cancelling out. For any other prior, one would have to model the selection, leading to the same final posterior on $H_0$. This is all under the assumption that no limits have been imposed on the distances, as is implied by the use of the maximum-likelihood distance values---no matter how large or small they are---in Eqs.~\ref{eq:post_H0_k} and~\ref{eq:zsel_chi2}.

When $\sigma_z=0$, volume-limited and redshift-selected samples are identical. This means that one can either treat them in the volume-selected way as discussed in the rest of the paper, or in the redshift-selected way as discussed here. For the former, one truncates the prior on the latent distances at $r_\text{min}$ and $r_\text{max}$. Since for $\sigma_z=0$ distances are deterministically related to $z$ given an $H_0$ sample, this means rejecting $H_0$ samples where any implied distance is outside the prior bounds. Using $k=2$ and no selection modelling then gives an unbiased result, but any other $k$ is biased. For the latter, one achieves the same result by not truncating the latent distances but rather multiplying the posteriors by $H_0^{1+k}$ (Eq.~\ref{eq:zsel}). Thus the uniform-in-$\mu$ prior without imposing any prior bounds on distances or any selection effects is unbiased, while the volume prior would require the posterior to be multiplied by $H_0^3$ per object. Indeed one could choose any value of $k$ in this case, since the $H_0^{(1+k)}$ factor would correct for it and achieve the same unbiased result.

Our treatment here would need to be generalised to allow for peculiar velocity modelling or inhomogeneous Malmquist bias, in which case one must also marginalise over the source sky position~\citep{CH0}. We have also assumed the absence of higher-order cosmographic terms and any dependence on sky position in the forward model. Including these effects would necessitate a more complex integral that is probably analytically intractable.

\subsection{Magnitude selection}\label{sec:sel_m}

If the selection is defined in apparent magnitude $m$, we instead have
\begin{equation}
\begin{split}
    p&(S=1\mid H_0)=\\
    &=
    \iiint \dd m\,\dd z\,\dd r\;
    p(S{=}1\mid m)\,
    \mathcal{L}(m\mid r)\,
    \mathcal{L}(z\mid H_0, r)\,
    \pi(r),
\end{split}
\end{equation}
where $p(S{=}1 \mid m)$ is the magnitude selection indicator. If the absolute magnitude is assumed known, $z$ enters only through the redshift likelihood. Since $\int \dd z\,\mathcal{L}(z\mid H_0, r) = 1$, the marginalisation over $z$ then removes the dependence on $H_0$, so the selection factor does not affect the posterior. Thus $k=2$ is required if the sources are intrinsically uniformly distributed in volume, the only difference to the volume-limited case being that one should not impose any $r_\text{min}$ or $r_\text{max}$ limits on the latent distances.

This crucially assumes that the distance indicator is pre-calibrated so that $M$ is effectively known. A first-principles distance ladder must infer this jointly with the other parameters. In this case the selection term depends on $M$, and affects $H_0$ due to the degeneracy between them. In eq.~31 of~\cite{CH0}, it is shown that combining some simplifying assumptions with the definition $\mu \equiv m - M$ implies that the selection term takes the form $p(S=1\mid M) \propto 10^{-3M/5}$. When both $M$ and $H_0$ are inferred jointly, they are positively degenerate: increasing $M$ implies fainter sources, which must then be closer to match the observed $m$ and hence require a higher $H_0$ to match the observed $z$. The selection term therefore drives $M$ towards higher (fainter) values, compensating for the preferential detection of brighter sources and consequently increasing the inferred $H_0$. This effect is however smaller than redshift selection which increases $H_0$ directly, such that the magnitude-selected $H_0$ result lies between the volume-limited and redshift-selected results~\citep{CH0}.

\subsection{Joint selection}\label{sec:sel_zm}

We can also consider the case of a joint redshift and magnitude selection. This produces
\begin{equation}
\begin{split}
    p(S=1\mid H_0)
    &=
    \iiint \dd m\,\dd z\,\dd r\; p(S=1\mid m)p(S=1\mid z)\\
    &\quad\times\mathcal{L}(m\mid r)\, \mathcal{L}(z\mid H_0, r)\, \pi(r).
\end{split}
\end{equation}
In this case, a simple change of variables cannot be used to extract the factors of $H_0$ from the integral, as it also appears explicitly in the magnitude likelihood. Furthermore, since $H_0$ is in the integral, the inference becomes explicitly sensitive to the functional forms of both selection terms, which must therefore be known. More constructively, if the selection function is e.g. a Heaviside step function, the integrations over $m$ and $z$ convert the Gaussian likelihoods into Gaussian cumulative distribution functions or error functions, reducing the selection term to a one-dimensional integral over $r$.

We can distinguish two asymptotic regimes as a function of $H_0$. At small $H_0$, the distance prior dominates the selection: all sources have low redshifts and pass the redshift selection cut, making the selection term effectively independent of $H_0$ and hence retaining the volume-limited result. At large $H_0$, the redshift selection limits the sample: all sources satisfy the magnitude cut but have high observed redshifts, so the selection term scales as $1/H_0^3$, as in the case of redshift selection only. In the intermediate regime, the slope smoothly transitions between these two limits. The exact point at which this transition occurs will depend on the sample, but we can at least say that, as with magnitude selection, the joint-selection $H_0$ result will be between the no-selection and redshift-selection results.

\subsection{Phenomenological selection model}\label{sec:sel_phenomenological}

\citet{Lavaux_2015} proposed a phenomenological model for selection that alters the distance prior rather than introducing extra factors into the posterior to account for the impact of missing objects. This is given by
\begin{equation}\label{eq:sel_phen}
    \pi(r) = Z \: r^p \: \exp(-(r/R)^q),
\end{equation}
with normalising proportionality constant
\begin{equation}
    Z \equiv \frac{q}{R^{1+p} \: \Gamma\left(\frac{p+1}{q}\right)}.
\end{equation}
This is a power-law rise at low $r$ (encompassing the volume prior $\pi(r) \propto r^2$ as a special case) followed by an exponential decay with onset distance and steepness set by two further free parameters. $p$, $R$, and $q$ are then inferred jointly with any other parameters (we adopt wide uniform priors on them). By favouring smaller distances beyond the mode of the distribution, this prior favours larger $H_0$ and thus mimics the effect of explicit selection modelling. However, it does not model the selection function in a principled manner, and is therefore not to be preferred for precision inference.

\subsection{Biases of different methods under different selection types}\label{sec:full_bias_results}

\begin{figure*}
    \centering
    \includegraphics[width=\textwidth]{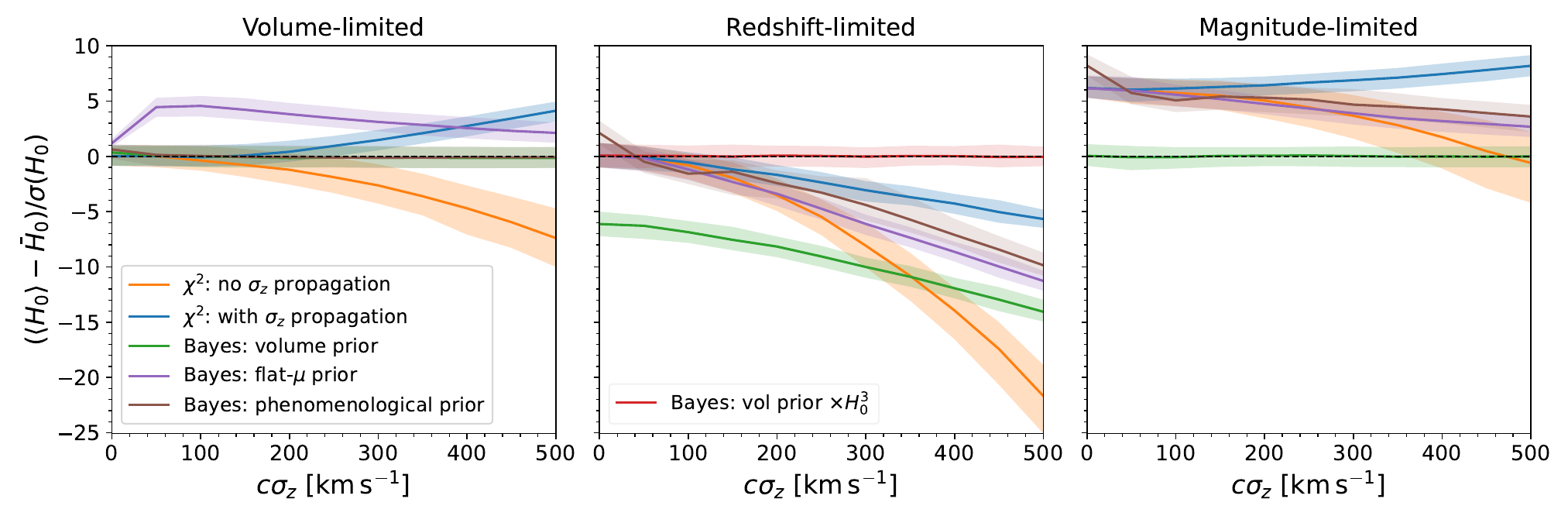}
    \caption{Relative bias $\mathcal{B}(H_0)$ in the inferred Hubble constant with the Bayesian forward model and $\chi^2$ estimator for the fiducial distance-ladder setup, shown here as a function of $\sigma_z$ separately for volume-, redshift-, and magnitude-limited samples. The lines show the median values over 200 mock datasets at each $\sigma_z$, while shaded bands show the 1$\sigma$ range. We show results for the uniform-in-volume and uniform-in-$\mu$ distance priors for the Bayesian method with and without propagating the redshift uncertainty (second term on the right hand side of Eq.~\ref{eq:chi2_sigmaz_propagation}) for the frequentist method. For the Bayesian method, the uniform-in-volume prior produces the correct result for volume- and magnitude-limited samples, and also for redshift selection once the appropriate $H_0^{3N}$ factor is included (Sec.~\ref{sec:sel_z}). The uniform-in-$\mu$ prior is only unbiased (by chance) in the case of redshift selection as $\sigma_z\rightarrow0$. For $\chi^2$, the volume-limited case follows the analytic expectation from Eq.~\ref{eq:chi2_bias_volume} if $\sigma_z$ is not included, but reverses sign when it is.
    The method is similarly biased for the magnitude- and redshift-selected samples. The phenomenological model manages to recover the volume prior in the case of volume-limited selection, but is otherwise biased.
    }
    \label{fig:bias_chi2}
\end{figure*}

Fig.~\ref{fig:bias_chi2} shows $\mathcal{B}(H_0)$ as a function of $c\sigma_z$ for volume-, redshift- and magnitude-limited samples for each of the methods described above. These tests use the same model parameters as in Sec.~\ref{sec:mock}, except that for the redshift-limited sample we impose a maximum observed redshift of $\bar{H}_0 r_{\max}$, while for the magnitude-limited sample we impose a maximum apparent magnitude of $\mu(r_{\max}) + M$, where $r_{\max} = 100~\mathrm{Mpc}$. In both cases, we draw true distances extending well beyond $r_{\max}$ and apply rejection sampling to retain $N_{\mathrm{gal}} = 2000$ galaxies within the selection limits. We create 200 independent mock datasets at 11 equally-spaced $c\sigma_z$ values between 0 and 500 km/s inclusive, showing the median $\mathcal{B}(H_0)$ with solid lines and the 16$^\text{th}$-84$^\text{th}$ percentile range with bands. It is important to remember that $c\sigma_z$ includes a contribution from peculiar velocity uncertainties, which introduces a scatter between predicted cosmological redshift and observed redshift. Thus even if observed redshifts are arbitrarily well-measured, $c\sigma_z \approx 300$ km/s in the absence of a peculiar velocity model, and $\approx150-200$ km/s with one~\citep{VFO}.

Fig.~\ref{fig:bias_chi2} validates the above results: the volume prior is naturally unbiased for volume-limited and magnitude-selected samples (assuming that $M$ is known), while for the redshift-selected case, it is readily made unbiased by including the selection factor $H_0^{3N}$ in the posterior. The uniform-in-$\mu$ prior is biased in all cases except for the case of redshift selection and $\sigma_z=0$, where the volume prior serendipitously cancels with the selection factor. The phenomenological model recovers the volume prior for volume-limited data by putting the exponential truncation beyond the known upper limit $r_\text{max}$ and fitting $p\approx2$, but does not achieve the correct result for the other types of selection: it remains roughly as biased as the other biased models.

Turning now to the $\chi^2$ results, we see first that this estimator is unbiased for a volume-limited sample when $\sigma_z = 0$. This follows from the equivalence of volume-limited and redshift-selected samples in that case and the fact that the redshift selection correction vanishes when $k=-1$, as is effectively assumed by that estimator. Ironically, neglecting \emph{both} the volume prior \emph{and} the selection effect (truncation of distances) produces the right answer in this case. It does however become biased for any $\sigma_z>0$, causing the volume-limited and redshift-selected cases to come apart. Neglecting $\sigma_z$ propagation yields a negative bias consistent with Eq.~\ref{eq:chi2_bias_volume}. Propagating $\sigma_z$ using Eq.~\ref{eq:chi2_sigmaz_propagation} reverses the sign, producing a bias of similar magnitude but positive.

If however the sample really is redshift-selected, then---provided the $\sigma_z$ uncertainty is propagated---it will continue to be largely unbiased for $H_0$ because the $H_0^{3N}$ redshift selection effectively converts the correct $k=2$ prior to $k=-1$. In other words, when both the distance prior and redshift selection are modelled self-consistently (under the simplifying assumptions discussed in Sec.~\ref{sec:sel_z}), they become equivalent to the $\chi^2$ estimator for any choice of $k$, owing to the fortuitous cancellation of the $H_0$ dependence.
The magnitude-limited sample shows a positive bias in $H_0$, which decreases when $\sigma_z$ is ignored but grows when $\sigma_z$ is propagated.
The $\chi^2$ estimator is therefore only approximately valid for volume- and redshift-selected samples; if the selection is instead in apparent magnitude, it will strongly overestimate $H_0$ even for $\sigma_z\rightarrow 0$. This is because the magnitude-selection correction is independent of $H_0$ when $M$ is known (Sec.~\ref{sec:sel_m}), so it cannot cancel the incorrect distance prior. In that case, one must adopt the volume prior.
Note that the exact numerical values of the bias depend on the specifics of the problem, such as the underlying true distance distribution; if expressed as a relative bias in units of the standard deviation, they would also depend on $N$ and $\sigma_m$.

Our overall conclusion is that the \emph{only} way to achieve the correct result (without adding on a ``bias correction'' tailor-made to remove bias) is to adopt the volume prior and account for selection in the principled manner of Sec.~\ref{sec:selection}.

\section{Case study I: CosmicFlows-4}\label{sec:cf4}

Having seen that the distance prior and selection effects can have a significant impact in principle, we now wish to know how much difference they make for real-world distance-ladder inferences of $H_0$. For our first case study, we investigate the CosmicFlows-4 (CF4) dataset, which is a compilation of 55,874 individual galaxy distances out to $z\approx0.1$, the largest of its kind~\citep{CF4}. The distances are derived from a variety of indicators, including the Tully--Fisher relation (TFR), Fundamental Plane (FP), Type Ia supernovae (SNe~Ia), and surface brightness fluctuations (SBF). These have been pre-calibrated so that the database quotes only the CMB-frame velocity $cz$ and the distance moduli, along with their (assumed independent) uncertainties.

\subsection{Volume prior}\label{sec:cf4_prior}

Have the distance moduli quoted in the CF4 catalogue already had the volume prior applied? \citet{CF4} mention in their sec.~4.2 that ``larger proposed distances for each galaxy end up being up-weighted to account for the increased cosmological volume in which a galaxy could be found''. \citet{Springob} and~\citet{Howlett}, providing subcatalogues for CF4, attempt to implement this with a parameter $f_n$ that is claimed to account for the volume effect. However it is simply the integral of a Gaussian likelihood with a flux limit to account for selection, such that when selection is unimportant (left side of fig.~9 of~\citeauthor{Howlett} and fig.~5 of~\citeauthor{Springob}), $f_n\rightarrow1$ and there is no prior preference for larger distances. Indeed it is stated explicitly around eq.~18 in~\citeauthor{Springob} that $f_n$ drops out in the case of a volume-limited survey. It therefore appears that the volume prior is not included and $f_n$ only models selection effects.

This is corroborated in other CF4-related papers. \citeauthor{Howlett} defines a log-distance ratio $\eta \equiv \log_{10}(d_\text{FP}/d_z)$ (up to a group correction), where $d_\text{FP}$ is the maximum-likelihood distance from the FP and $d_z$ is the redshift distance. $\eta$ is then given a uniform prior, which corresponds to a flat prior on $\mu$ nearby where peculiar velocities dominate, but an even more steeply declining prior ($\appropto r^{-2}$) further out in the Hubble flow. The TFR subcatalogue analysis papers~\citep{Kourkchi_0,Kourkchi_1,Kourkchi_2} make no mention of the effect, while the recent ``prior-free'' reanalysis of~\citet{Duangchan_2025} explicitly neglects the volume term in their eqs.~7-8. The CosmicFlows-2 paper~\citep{CF2} is the last place one can find a clear statement: ``we make no adjustments for the distribution Malmquist effects in our reported distances''.

To implement the volume prior, we focus on the TFR subset of the data, impose $cz > 4000~\kmsec$ to limit the impact of peculiar velocities, and fix the deceleration parameter $q_0$ to $-0.595$ and the jerk parameter $j_0$ to 1 \citep[following][]{CF4}. This leaves 8,951 measurements. We take the distance moduli directly from the catalogue, implicitly adopting Eq.~\ref{eq:simple_nlp} with $k=-1$, $\sigma_z\rightarrow 0$ and no bounds on the distances corresponding to a volume limit. We compare these to the predicted distance moduli
\begin{eqnarray}
	\mu_i &=& 5 \log_{10}(\frac{cz_i}{H_0} f) + 25, ~~\text{where}\\
    \label{eqn:f}
	f &\equiv& 1 + \frac{1-q_0}{2} \: z - \frac{1 - q_0 - 3 q_0^2 + j_0}{6} \: z^2,
\end{eqnarray}
We thereby infer $H_0 = 75.13 \pm 0.16~\kmsecMpc$, in agreement with~\citet{CF4} (who likewise neglect the inhomogeneous Malmquist bias that may have a non-negligible impact at such low redshift).\footnote{The methodology is however different to that used in~\citet{CF4}. They compute for each galaxy $\log_{10} H_i = f_i c z_i/d_i$ (with $f$ given by Eq.~\ref{eqn:f}) and then report the mean and uncertainties from this set of values. We instead take all galaxies into account simultaneously in a global $H_0$ fit.} Imposing instead $k=2$ (shifting the distance moduli according to Eq.~\ref{eq:simple_muhat} using the per-object $\sigma_m = \sigma_\mu$), we instead find $H_0 = 66.85 \pm 0.14~\kmsecMpc$. This is a dramatic $\approx 55\sigma$ shift with respect to the $k = -1$ result. It reflects the relatively large uncertainties on the TFR distance moduli in CF4, ranging from 0.28 to 0.8 with a median value of 0.41, combined with the large sample size producing a very precise estimate of $H_0$. $\langle \sigma_\mu^{-2} \rangle$ is 5.45, so Eq.~\ref{eq:simple_deltaH0_varerr} predicts $\Delta \ln \hat{H}_0 = 0.12$, which translates to $\Delta \hat{H}_0 \approx 8~\kmsecMpc$, as measured. Similar results are obtained using any other subset of galaxy distances in CF4, the full set, or the group catalogue.

This shift is illustrative only; we are not arguing that CF4 implies $H_0 = 66.85 \pm 0.14~\kmsecMpc$. This is mainly because this analysis neglects various important sources of systematic uncertainty, which according to~\citet{CF4} contribute $\approx 3~\kmsecMpc$ and hence dominate the overall error budget. Chief among these is the covariance between the distance parameters (due to uncertainties in the inferred scaling relation parameters), which are treated here as uncorrelated Gaussian measurements. A more rigorous reanalysis would go back to the raw observables and infer the full set of distances simultaneously with cosmological and nuisance parameters. It is also critical to model selection effects arising from e.g. flux or redshift limits, as we discuss next.

\subsection{Selection effects}\label{sec:cf4_selection}

In a volume-limited setting, the baseline fitting method would correspond to $\chi^2$ rather than the Bayesian flat-$\mu$-prior method which would include bounds on the distances. This means that if CF4 were truly volume-limited, the baseline method would return the correct $H_0$ ($\approx75$ km/s/Mpc), which could alternatively be recovered by implementing both the volume prior and the selection effect (bounds on distances). This, and the results below, follow the $\sigma_z\rightarrow0$ results of Fig.~\ref{fig:bias_chi2}. However, the sample is clearly not volume-limited:
although the observed redshift distribution approximately follows $z^2$ at low $z$, it is truncated because more distant sources are not observed. The $V_\text{CMB} > 4000~\kmsec$ cut used above imposes a redshift selection but makes little difference: without it $H_0 =74.80 \pm 0.13~\kmsecMpc$ for $k=-1$ and $66.43 \pm 0.12~\kmsecMpc$ for $k=2$. The entire sample selection however makes a significant difference, as more distant galaxies are clearly preferentially excluded.

\subsubsection{Redshift selection}

Sec.~\ref{sec:sel_z} shows that the multiplicative correcting factor for the case of redshift selection exactly cancels the effect of a power-law distance prior regardless of the value of $k$, and is unity for $k=-1$. The upshot is that the uniform-in-$\mu$ prior implicitly assumed by CF4's maximum-likelihood fit is correct in this case, so that $H_0 \approx 75~\kmsecMpc$. This is the same equivalence between redshift selection and volume selection in the $\sigma_z\rightarrow0$ limit that we saw before.

\subsubsection{Magnitude selection}

Since the absolute magnitudes of the standard candles have been pre-calibrated in constructing the CF4 catalogue (allowing distances to be effectively treated as observables), Sec.~\ref{sec:sel_m} shows that magnitude selection would not impact $H_0$ when inferred using the volume prior. Were this the case, the correct $H_0$ would therefore be $\approx 67~\kmsecMpc$. If the zero-point ($M$) was also inferred, neglecting the selection term would prevent the model from recognising that the sample preferentially includes brighter sources. As a result, $M$ would be inferred to be too bright, placing all sources at systematically larger distances and, consequently, $H_0$ would be between 67 and $75~\kmsecMpc$. This would also be the case if joint magnitude and redshift selection were in operation. (Of course we are assuming that the pre-calibrated distances are reliable; were these calibrations also to contain faulty assumptions concerning priors, selection or inference methodology the $H_0$ result could change further.)

\subsubsection{Phenomenological selection}

\begin{figure*}
  \centering
  \begin{subfigure}[t]{0.44\textwidth}
    \centering
    \includegraphics[width=\linewidth]{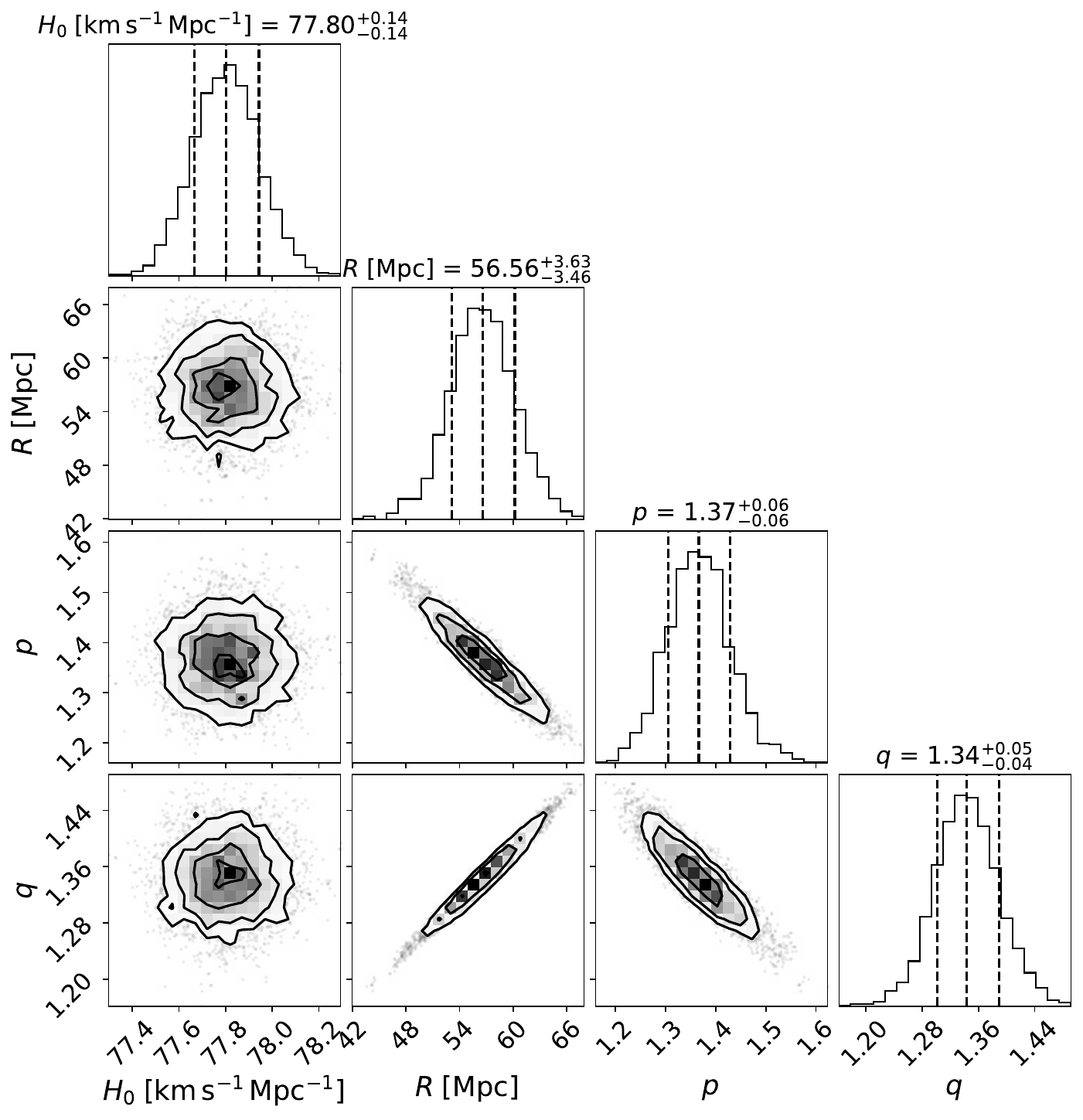}
  \end{subfigure}
  \hfill
  \begin{subfigure}[t]{0.55\textwidth}
    \centering
    \includegraphics[width=\linewidth]{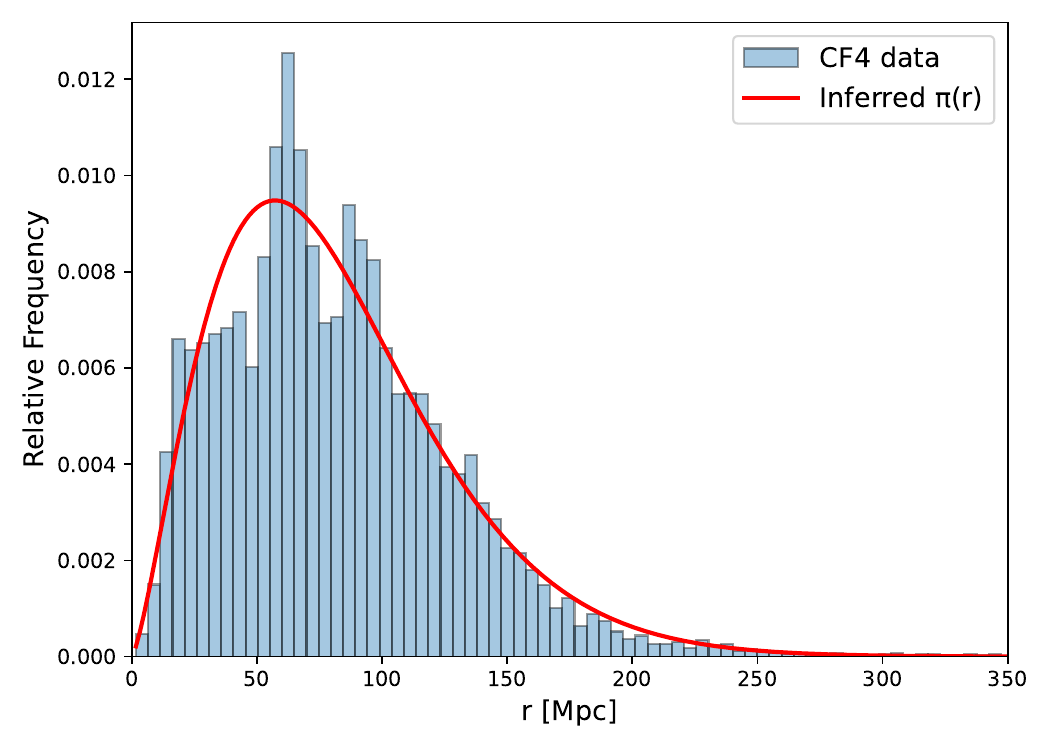}
  \end{subfigure}
  \caption{Corner plot (left) and distance histogram with overlaid best-fit prior (right) for the phenomenological selection model applied to the Tully--Fisher CF4 data without $4000~\kmsec$ cut.}
  \label{fig:cf4_phen}
\end{figure*}

Applying the method of Sec.~\ref{sec:sel_phenomenological} rather than any power-law distance prior, we infer $H_0 = 78.10 \pm 0.16~\kmsecMpc$ for the Tully--Fisher CF4 sample with $4000~\kmsec$ redshift cut. The results are similar when the redshift cut is not applied, in which case the distance distribution follows more closely the functional form assumed by this selection model; the full corner plot and distance distribution with best-fit prior overlaid are shown for this case in Fig.~\ref{fig:cf4_phen}. $H_0$ is increased because this pushes distances down, which increases their prior probabilities if they are beyond the peak of Eq.~\ref{eq:sel_phen} as most of them are. The result is however likely to be sensitive to the priors on $R$, $p$, and $q$ (which have been set to uniform without good justification) as this determines what shapes are preferred for the distance prior. This model is in any case indicative only, as it does not model selection in any principled fashion. Indeed we see from Fig.~\ref{fig:bias_chi2} that this method is clearly biased in general. It is however the only fully-specified model available for this dataset, as implementing the principled modelling would require a much more detailed knowledge of the selection effects impacting the sample, made much more challenging by the fact that the sample is composed of multiple quite different sub-catalogues. We leave quantification of this for a future, more thorough recalibration of the CF4 distance ladder.

\section{Case Study II: SH0ES}\label{sec:sh0es}

The most precise distance-ladder measurement of $H_0$ is from the SH0ES collaboration, who construct a three-rung ladder comprising geometric anchors, Cepheids, and SNe~Ia \citep{Breuval_2024}. Given the importance of this result to the Hubble tension, it is necessary to know whether this inference is unbiased in relation to the distance prior and selection effects.

By treating distances as observables, CF4 effectively worked in the limit $\sigma_z\rightarrow 0$, validating the simplified posterior of Eq.~\ref{eq:simple_nlp}. In this case, we know that $k=-1$ works only for redshift-selected samples, but otherwise it overestimates $H_0$. The situation is more complex for SH0ES because it does not assume $\sigma_z=0$. We assume that the basic inference method of SH0ES is the $\chi^2$ estimator, including propagation of the $\sigma_z$ uncertainty. For realistic $c\sigma_z \approx 300$ km/s, Fig.~\ref{fig:bias_chi2} shows that this overestimates $H_0$ for volume- or magnitude-selected samples but underestimates it for redshift-selected samples.

SH0ES employs a method for ``bias-correcting'' SN apparent magnitudes for selection effects by comparing with idealised simulations, which is effectively designed to ``make up the difference'' with the true $H_0$ in Fig.~\ref{fig:bias_chi2}.
Fig.~3 of~\citet{Kessler_Scolnic} and~\citet{Popovic} show that these bias corrections increase distance moduli and hence reduce $H_0$ (note that the convention in those papers was that the distance shifts $\Delta\mu$ were \emph{subtracted}). This could plausibly produce an unbiased inference for either a redshift- or magnitude-selected sample. However, it must be borne in mind that SN selection is more complicated than our models assume: SNe are detected based on individual ``epoch'' magnitudes at one point in time before standardisation. This depends on the noise realisation, which is a function not only of magnitude but also of colour. Besides discovery, there is also a selection for spectroscopic follow-up to identify the SN type. The conjunction of these effects necessitates a simulation-based approach which is beyond our scope; we are only able to check that the SH0ES pipeline could plausibly effectively account for both the volume effect and selection based on our limiting cases.

While it is beyond the scope of the paper to model selection in SH0ES using the formalism of Sec.~\ref{sec:selection}, it is instructive to quantify the volume prior in a SH0ES-type setup to ascertain the magnitude of its effect on $H_0$. Note that this is illustrative only: not only does it neglect the crucial impact of selection, but the frequentist method is also not equivalent to the Bayesian flat-$\mu$-prior method, which for $\sigma_z>0$ would not be fully fixed by adding the volume prior alone---regardless of the selection.

We download the publicly available data\footnote{\url{https://github.com/PantheonPlusSH0ES/DataRelease/tree/main/SH0ES_Data}} and follow the prescription in sec. 2 of~\citet{Riess_2022}. Without any modifications, we find $H_0 = 73.04 \pm 1.02~\kmsecMpc$, in near-perfect agreement with the value quoted in the paper. To implement the volume prior in the Cepheid host distances, which are free parameters in the inference, we add $3 \alpha^{-1} \sum_i \mu_i$ to the log-posterior (Eq.~\ref{eq:simple_nlp}), where $\mu_i$ are the sampled host distance moduli. This lowers $H_0$ to $72.01~\kmsecMpc$ without changing the uncertainty. This is a 1.4 per cent or $1\sigma$ shift, with Cepheid distance moduli increased by 0.03 on average.

Modifying the Hubble-flow SN distances for the volume prior requires a different approach because these are not considered free parameters in the SH0ES inference, so we do not have sampled distance moduli. This may be achieved by increasing the SN apparent magnitudes by an amount that emulates the effect on $H_0$ of the increased distances caused by $k=-1 \rightarrow k=2$. We begin with the generalisation of Eq.~\ref{eq:simple_nlp} that includes a covariance linking the measured magnitudes:
\begin{equation}\label{eq:nlp_cov}
-\ln P(\boldsymbol\mu)
=\tfrac12(\boldsymbol\mu-\boldsymbol m+M\mathbf 1)^{T}C^{-1}(\boldsymbol\mu-\boldsymbol m+M\mathbf 1)
-\frac{(k+1)}{\alpha}\,\mathbf1^{T}\boldsymbol\mu,
\end{equation}
where \(\mathbf1\) is the vector of ones and \(C\) is the covariance matrix. The SH0ES data vector contains \(\boldsymbol{y}\equiv\boldsymbol{m} - \alpha\ln(c\boldsymbol{z})\)
(dropping the ``--25'' in accordance with our magnitude definition). In the limit of small \(\sigma_z\), we have
\(\boldsymbol\mu = \alpha\ln(c\boldsymbol{z}) - \alpha\ln H_0\,\mathbf1\),
which lets us rewrite Eq.~\ref{eq:nlp_cov} in terms of \(H_0\):
\begin{equation}\label{eq:nlp_cov_H0}
    \begin{split}
    -\ln P(H_0)
=&\tfrac12(M\mathbf{1} - \alpha\ln H_0\,\mathbf{1} -\boldsymbol{y})^{T}
      C^{-1}(M\mathbf{1} - \alpha\ln H_0\,\mathbf{1} -\boldsymbol{y})\\
&-(k+1)\,[\,\mathbf1^{T}\ln(c\boldsymbol{z}) - N\ln H_0\,],
    \end{split}
\end{equation}
where \(N\) is the length of the data vector.

Differentiating with respect to \(\ln H_0\) gives
\begin{equation}
\frac{\partial(-\ln P)}{\partial\ln{H_0}}
= -\alpha\,\mathbf1^{T}C^{-1}(M\mathbf{1} - \alpha\ln H_0\,\mathbf{1} -\boldsymbol{y})
  + (k+1)N.
\end{equation}
Setting this to zero gives the equation for the MAP \(H_0\):
\begin{equation}
    \alpha\,\mathbf1^{T}C^{-1}(M\mathbf{1} - \alpha\ln \hat{H}_0\,\mathbf{1} -\boldsymbol{y})
    = (k+1)N.
\end{equation}
This implies
\begin{equation}
    \ln\hat{H}_0
    = \frac{\tfrac{1}{\alpha}\,\mathbf1^{T}C^{-1}(M\mathbf{1}-\boldsymbol{y})
      - \tfrac{(k+1)N}{\alpha^{2}}}
      {\mathbf1^{T}C^{-1}\mathbf1}.
\end{equation}
To emulate the shift in \(H_0\) produced by \(k_1 \rightarrow k_2\)
through a modification \(\boldsymbol{y}\rightarrow\boldsymbol{y}'\), we require
\begin{equation}
    \ln\hat{H}_{0,k_1}(\boldsymbol y') = \ln\hat{H}_{0,k_2}(\boldsymbol y),
\end{equation}
which leads to the scalar condition
\begin{equation}
    \mathbf1^{T}C^{-1}\Delta\boldsymbol{y}
    = \frac{(k_2-k_1)N}{\alpha},
\end{equation}
where \(\Delta\boldsymbol{y}\equiv\boldsymbol{y}'-\boldsymbol{y}\).
The solutions to this equation are
\begin{equation}
    \Delta \boldsymbol{y} = \frac{(k_2-k_1)}{\alpha}\,C\,\mathbf1 + \boldsymbol{v},
\end{equation}
where $\mathbf{1}C^{-1}\boldsymbol{v} = 0$. Since $\boldsymbol{v}$ does not affect the inferred $\hat{H}_0$ (only shifting the data in directions irrelevant to that), we are free to choose $\boldsymbol{v}=\mathbf{0}$. This yields the component-form solution
\begin{equation}
    \Delta y_i = \frac{(k_2-k_1)}{\alpha}\,\sum_{j} C_{ij}.
\end{equation}
(Note that this derivation assumes small $\sigma_z$ and a linear Hubble expansion, allowing us to approximate that $r_i H_0 \approx c z_i$. These are reasonable approximations, but deviations from them will cause small alterations to the result in practice.) Implementing this in conjunction with the shift to the Cepheid hosts, we find an overall reduction of $H_0$ to $71.31~\kmsecMpc$. Implementing just the SN correction but not the Cepheid correction would give $H_0 = 72.30~\kmsecMpc$.

It is important to emphasize that we are not suggesting that these lower $H_0$ values are the correct result for SH0ES, which should already implicitly include the volume prior through their simulation-based bias corrections.
The complex selection effects at play---also modelled by the SH0ES bias correction scheme---will increase $H_0$ beyond a na\"ive no-selection expectation. This can be illustrated by reference to~\citet{HM}. Inspired by the first version of our paper, those authors implemented the volume prior across the SH0ES distance ladder, arguing that $H_0$ is therefore significantly lower than the nominal SH0ES value. However, aside from the issue of whether or not the SH0ES bias correction scheme implicitly corrects for homogeneous Malmquist bias, neglecting selection effects as~\citet{HM} do produces a generative model that is clearly discrepant with the data. This is illustrated explicitly for the first rung of the SH0ES distance ladder (Cepheids in the Milky Way) in~\citet{Stiskalek_MW}, where it is shown that---as here for the case of a redshift-limited sample with $\sigma_z\rightarrow0$---accounting for selection in conjunction with a physical prior restores approximate consistency with the $\chi^2$ method and hence the fiducial SH0ES results. A secondary issue is that the geometry of the Milky Way is not spherical, so the physical prior for Milky Way Cepheids is not $r^2$. A principled full analysis---begun in~\citet{CH0,Stiskalek_MW}---remains for future work.

\section{Discussion and Conclusion}\label{sec:conc}

We have shown that the prior used for galaxy distances in distance-ladder studies can have a significant impact on the inferred value of $H_0$. To illustrate this, we set up a simplified distance-ladder inference of $H_0$ neglecting redshift uncertainties and assuming a volume-limited sample. We then calculate analytically the shift to the best-fit $H_0$ (both directly in $\kmsecMpc$ and as a multiple of the $H_0$ uncertainty) between different choices of exponent for a power-law distance prior. To determine which prior gives an unbiased inference of $H_0$ and allow for arbitrary uncertainties, we also perform the inference numerically on mock data.

We find that $\pi(r)\propto r^2$ produces an unbiased posterior, while any other choice results in bias. This is a direct consequence of the assumption in the mock generation that objects are uniformly distributed in three-dimensional space, as in the real Universe. The issue is important because many state-of-the-art distance ladders currently do not impose $\pi(r) \propto r^2$ but rather, by maximising likelihoods for distance moduli, implicitly impose $\pi(r) \propto 1/r$. We show that \emph{unless the sample is strictly redshift-selected and redshift uncertainties are negligible},
this results in a bias, which in the case of volume- or magnitude-limited selection is low in distances and high in $H_0$. Since the magnitude of the $H_0$ shift in $\kmsecMpc$ is independent of the sample size, the relative bias in units of the width of the $H_0$ posterior scales with $\sqrt{N}$. Thus the relative bias (as a multiple of the uncertainty) will grow for larger future datasets.

Many of our calculations have assumed a volume-limited survey, allowing us to neglect selection effects. This is to demonstrate that the volume prior is entirely independent of selection, a point that is sometimes lost in the literature when both are called ``Malmquist bias''. While we have shown that selection effects can practically (partially) undo the effect of the volume prior, they are conceptually unrelated effects: the prior describes the \emph{intrinsic} distribution of sources, while selection affects which objects from the predicted population enter the sample, and hence the likelihood of the \emph{observed} data. They come together only on application of Bayes' Theorem. This is also why the phenomenological selection model which alters the distance prior is at best approximate.

Our more in-depth study is of the CF4 dataset, for which adopting an $r^2$ prior instead of the $1/r$ prior to which their maximum-likelihood analysis is equivalent
shifts the inferred $H_0$ down by $8.3~\kmsecMpc$ ($55\sigma$) to a best-fitting value of $H_0 = 66.9~\kmsecMpc$. Assuming the rest of the CF4 modelling is correct, this would hold for the case of
magnitude-limited selection. For redshift-selection (or a hypothetical volume-limited sample), the previously-reported value of ${\approx}\,75~\kmsecMpc$ would be correct. The truth is likely in between these limiting cases, although the fact that the phenomenological model of~\citet{Lavaux_2015} yields $78~\kmsecMpc$ may suggest that an $H_0$ value at the higher end is more realistic. We also investigate the SH0ES sample, for which we find the volume prior has a $1.7\sigma$ effect on $H_0$, likely already accounted for within the SH0ES pipeline.

The issue is easier to see in Bayesian (re)analyses of the distance ladder, which do (or at least should) treat distances as inferred parameters and hence adopt one of our Bayesian methods (rather than $\chi^2$). To our knowledge all such analyses fail to account for the volume effect:~\citet{March} implicitly uses a uniform prior on cosmological redshifts,~\citet{Feeney} and~\citet{Mandel_SNe} explicitly use uniform priors on distance moduli, and~\citet{Becker} and~\citet{Suvodip} explicitly use a uniform prior on distance. This latter produces a result between the uniform-in-$\mu$ and uniform-in-volume priors, and would therefore underestimate $H_0$ in the case of redshift selection but may approximately account for the effect of a joint redshift-and-magnitude selection.

Other Bayesian SN frameworks such as \texttt{UNITY}~\citep{UNITY} and \texttt{Steve}~\citep{Steve} do not treat distances as latent parameters at all. If working with distance moduli, this means they implicitly assume the uniform-in-$\mu$ prior, which again would require redshift selection in order to produce an unbiased $H_0$.
\texttt{CIGaRS}~\citep{CIGaRS} also does not forward-model the observables, but instead computes distance moduli deterministically from latent cosmological redshifts and the cosmological model, using a ``prior'' on redshifts from the distribution of host galaxies. \texttt{BayeSN}~\citep{BSN_1, Mandel_2011,BSN_2,Grayling_2024} adopts a uniform-in-$\mu$ prior both when fitting for the photometric distance modulus of an individual SN with a pre-trained model, and when using hierarchical Bayesian inference to train the model on a sample of SNe simultaneously to learn the population-level components of the spectral energy distribution (where this prior is multiplied by a distance--redshift likelihood constraint). Depending on the nature of the selection, the use of such incorrect priors could bias $H_0$ either up or down (Fig.~\ref{fig:bias_chi2}). Ultimately we see only two methods for unbiased distance ladder inference, one principled (a fully generative Bayesian forward model with physical distance prior and selection effects modelled from first principles) and one not (any old biased method with careful post-hoc corrections based on simulations in which both the intrinsic distribution of sources and applied selection cuts are accurate).

While we have focused on the impact on $H_0$, the volume prior also affects the inferred distances and anything derived from them (e.g. peculiar velocities). In both the large and small $\sigma_z$ limits, the MAP distances are given by (Eq.~\ref{eq:simple_muhat})
\begin{equation}
    \hat{r}_i(k) = \hat{r}_i(k=0) \: \exp\left(\frac{\sigma_m^2 (\ln 10)^2}{25} \: k\right).
\end{equation}
For $\sigma_m=0.1$ the coefficient of $k$ in the exponential is 0.0021, corresponding to a 0.64 per cent shift in $\hat{r}$ for $\Delta k = 3$. In contrast, for CF4 where $\langle \sigma_\mu^{-2}\rangle^{-1/2} = 0.43$ (Sec.~\ref{sec:cf4}) the effect is significantly larger and corresponds to a $\approx 12$~per cent increase in best-fit distances. This would cause a corresponding decrease in best-fit peculiar velocities, which could then impact inference of the growth rate of structure and the $S_8$ parameter. Since the magnitude of the effect scales inversely with the strength of the constraint (i.e. the relative importance of the likelihood and prior), the differential bias that it produces may cause two distance or peculiar velocity measures to appear discrepant when they are not, or vice versa. The same can be said of $H_0$ inferences: lower-precision measurements are biased high by a larger amount than higher-precision measurements. It is therefore crucial when comparing estimates of the distance~\citep{Antonio}, peculiar velocity~\citep{VFO}, and Hubble constant~\citep[e.g.][]{Review_Freedman,Review_Valentino,Review_Hu,Valentino_2025, H0DN_2025}.

Besides the requirement of a physical distance prior in Bayesian inference, our study highlights the vital need for accurate selection modelling. Some information on the nature of the selection can be found in the distribution of residuals between the inferred and predicted magnitudes; under magnitude selection this would be a function of redshift, but not under redshift selection. However, a principled accounting for selection effects requires an observational sample drawn from the parent population according to known, homogeneous criteria. Perhaps surprisingly this is rarely the case. Future distance-ladder samples should prioritise this, which will become easier with current and upcoming surveys with high completeness in the local Universe (e.g., for SNe~Ia, the Zwicky Transient Facility;~\citealt{ZTF_1, ZTF_2}). Otherwise a potentially significant systematic uncertainty from selection must remain in inferred parameters such as $H_0$. Further work is also needed to generalise the principled selection modelling to more realistic cases such as sky-dependence ~\citep[treated in][]{CH0} and, at high redshift, the inclusion of higher-order cosmographic terms and source evolution effects.

\section*{Acknowledgements}

HD, JAN, and IB are supported by Royal Society University Research Fellowship 211046. RS is supported by STFC Grant No. ST/X508664/1 and the Snell Exhibition of Balliol College, Oxford. We thank Matthew Colless, H{\'e}l{\`e}ne Courtois, Sebastian von Hausegger, Alan Heavens, Cullan Howlett, Mike Hudson, Guilhem Lavaux, Kaisey Mandel, Daniel Mortlock, Adam Riess, Daniel Scolnic and Aur\'elien Valade for useful discussions.

\section*{Data availability}

The CF4 data is publicly available at \url{https://edd.ifa.hawaii.edu/dfirst.php}, and the SH0ES data at \url{https://github.com/PantheonPlusSH0ES/DataRelease/tree/main/SH0ES_Data}. Our code is publicly available on GitHub \githublink.

\bibliographystyle{mnras}
\bibliography{Prior_impact_bbl}

\begin{appendix}

\section{More detailed analytic calculations of the distance prior effect}\label{sec:appendix}

The derivation of Sec.~\ref{sec:analytic} may be unsatisfactory to the mathematically minded reader. For such readers we provide here a more rigorous derivation, which also calculates $\hat{H}_0$ explicitly (not just its variation with $k$) and solves the opposite limit in which redshift uncertainties dominate.

From Eq.~\ref{eq:nlp_full} we find that $\hat{H}_0$ and $\hat{r}_i$---for arbitrary $\sigma_z$---satisfy
\begin{equation}\label{eq:MAPs_H0}
    \frac{\partial (-\ln\mathcal{P})}{\partial H_0} = \sum_{i=1}^N \frac{r_i}{c \sigma_z^2} \left(\frac{H_0 r_i}{c} - z_i\right) = 0.
\end{equation}
This implies that
\begin{eqnarray}
    \hat{H}_0 &=& \frac{c \sum_{i=1}^N z_i r_i}{\sum_{i=1}^N r_i^2}, \\
    \frac{\partial (-\ln\mathcal{P})}{\partial r_i} &=& \frac{\alpha}{\sigma_m^2 r_i} (M + \alpha \ln r_i - m_i) \nonumber \\
    &+& \frac{H_0}{c \sigma_z^2} \left(\frac{H_0 r_i}{c} - z_i\right) - \frac{k}{r_i} = 0.
\label{eq:MAPs_r}
\end{eqnarray}
The equations for $\{\hat{r}_i\}$ are coupled through $\hat{H}_0$ and generally require numerical solution, although further analytic progress may be made under the assumption that either the redshift or magnitude term dominates. Noting that residuals of $(M + \alpha \ln r_i - m_i)$ are expected to be $\mathcal{O}(\sigma_m)$ while residuals of $(H_0 r_i/c - z_i)$ are expected to be $\mathcal{O}(\sigma_z)$, this is the case if $c \sigma_z/H_0$ is either much larger or much smaller than $r \sigma_m/\alpha$. (This provides another way of justifying Eq.~\ref{eq:small_sigmaz} as the magnitude-uncertainty-dominated case.)
Here we explore both limits in more detail.

\subsection{The small-redshift-uncertainty limit}\label{sec:app_detailed}

Under Eq.~\ref{eq:small_sigmaz}, the second term on the right hand side of Eq.~\ref{eq:nlp_full} is replaced by the $\delta$-function constraint
\begin{align}\label{eq:delta}
    \delta\!\left(z_i - \frac{H_0 r_i}{c}\right)
    &= \frac{c}{H_0}\,
       \delta\!\left(r_i - \frac{c z_i}{H_0}\right),
\end{align}
so that the distances follow directly from the (assumed perfectly-known) redshifts given a model $H_0$. Minimising the negative log–posterior with respect to $\ln H_{0}$, while enforcing \(r_{i}=c z_{i}/H_{0}\) in the small-$\sigma_{z}$ limit, therefore introduces a Jacobian factor $c/H_0$ for each object. Substituting \(\ln \hat{r}_i = \ln(c z_i) - \ln \hat{H}_0\) into Eq.~\ref{eq:nlp_full} and \(B_i\equiv M+\alpha\ln(c z_i)-m_i\) while including this factor, we get that
\begin{equation}\label{eq:Nx}
-\ln \mathcal{P}(\ln{H_0})
 = \sum_{i=1}^{N}\frac{\big[B_{i}-\alpha \ln{H_0}\big]^{2}}{2\sigma_{m}^{2}}
   -\sum_{i=1}^{N} k \ln (c z_{i})
   + (k+1)N\ln{H_0} ,
\end{equation}
where \(B_{i}\equiv M+\alpha\ln (c z_{i})-m_{i}\). Differentiating and setting to zero yields
\begin{equation}
\alpha^{2} N \ln{\hat{H_0}}
 = \alpha \sum_{i=1}^{N} B_{i}
   - (k+1) N \sigma_{m}^{2},
\end{equation}
which implies
\begin{equation}
\ln \widehat{H}_{0}
   = \frac{1}{\alpha N}\sum_{i=1}^{N} B_{i}
     - \frac{(k+1)\sigma_{m}^{2}}{\alpha^{2}}
\end{equation}
Expanding $B$, this produces
\begin{equation}
\widehat{H}_{0}
   = \exp\!\left[
       \frac{M-\langle m\rangle}{\alpha}
       + \big\langle \ln (c z) \big\rangle
       - \frac{(k+1)\sigma_{m}^{2}}{\alpha^{2}}
     \right]
\end{equation}
where $\langle\cdot\rangle$ denotes the mean over the $N$ objects.

We can express this in units of the width of the $H_0$ posterior by finding the curvature at the MAP point from Eq.~\ref{eq:Nx}:
\begin{equation}
	\frac{\partial^2(-\ln\mathcal{P})}{\partial (\ln{H_0})^2}=\frac{\alpha^2 N}{\sigma_m^2}.
\end{equation}
This implies a variance of
\begin{equation}
	\sigma^2(\ln H_0)\simeq \frac{\sigma_m^2}{\alpha^2 N}.
\end{equation}
Propagating to \(\hat{H}_0\) gives
\begin{equation}\label{eq:sigvar}
	\sigma(H_0)\simeq \hat{H}_0\,\frac{\sigma_m}{\alpha\sqrt{N}}.
\end{equation}
From the expression for $\ln \hat{H}_0$, we have
\begin{equation}
	\hat{H}_0(k)=\hat{H}_0(k = 0)\,\exp \left(-k \alpha_m \right),
\end{equation}
where $\alpha_m \equiv \sigma_m^2/\alpha^2$. Thus for $k_2=k_1+\Delta k$ and defining $\Delta \hat{H}_0 \equiv \hat{H}_0(k_2) - \hat{H}_0(k_1)$, we find that
\begin{equation}\label{eq:deltaH0_app}
	\Delta \hat{H}_0 = \hat{H}_0(k_1)\left[ \exp \left( -\alpha_m \Delta k \right) - 1\right] \approx -\alpha_m\,\hat{H}_0(k_1)\,\Delta k,
\end{equation}
where the latter approximation holds for small $\alpha_m \Delta k$. Using Eq.~\ref{eq:sigvar}, the relative bias in units of the posterior standard deviation is therefore
\begin{equation}
	\frac{\Delta \hat{H}_0}{\sigma(H_0(k_1))}\simeq
	-\sqrt{N}\,\frac{\sigma_m}{\alpha}\,\Delta k,
\end{equation}
in agreement with Eq.~\ref{eq:simple_deltaH0rel}.

\subsection{The small-magnitude-uncertainty limit}\label{sec:app_other}

Here we investigate the opposite limit to Eq.~\ref{eq:small_sigmaz}, namely $\frac{c \sigma_z}{H_0} \gg \frac{r \sigma_m}{\alpha}$, such that the distance information is essentially coming solely from the distance indicator, with the redshift playing little role. In this case, the term containing $H_0$ in Eq.~\ref{eq:MAPs_r} can be neglected, producing
\begin{equation}\label{eq:H0_MAP}
\hat{r}_i \approx \exp \left( \alpha^{-1}(m_i - M) + k \alpha_m \right).
\end{equation}
We define $r_{i0} \equiv 10^{(m_i-M)/5}$ (MAP $r_i$ for $k=0$), so that
\begin{equation}
\hat{r}_i(k) \simeq r_{i0} \, \exp \left( k \alpha_m \right).
\end{equation}
Note that $k=0$ then corresponds to a distance that exactly produces the apparent magnitude from the true absolute one (i.e. maximises the likelihood), as this is the only constraint. This does \emph{not} however make the choice $k=0$ unbiased: the true distance is likely to be larger than the maximum-likelihood one due to the fact that more volume exists at higher $r$.

For a realistic magnitude uncertainty of $\sigma_m \approx 0.1$, $\alpha_m \approx 0.002$. Therefore for two values of $k$ separated by $\mathcal{O}(1)$, we can safely expand $\hat{H}_0(k)$ to first order in $\alpha_m \Delta k \equiv \alpha_m \left( k_2 - k_1 \right)$. Including explicit $k$-dependence where relevant, this yields
\begin{equation}
\hat{r}_i \left( k_2 \right) = \hat{r}_i \left( k_1 \right) \, \exp \left( \alpha_m \Delta k \right) \simeq \hat{r}_i \left( k_1 \right) \left( 1 + \alpha_m \Delta k \right).
\end{equation}
This lets us calculate
\begin{align}
\sum_i z_i \hat{r}_i(k_2) &\simeq \sum_i z_i \hat{r}_i(k_1) (1 + \alpha_m \Delta k) \nonumber \\
&= \sum_i z_i \hat{r}_i(k_1) + \alpha_m \Delta k \sum_i z_i \hat{r}_i(k_1),
\end{align}
\begin{align}
\sum_i \big(\hat{r}_i(k_2)\big)^2 &\simeq \sum_i \big(\hat{r}_i(k_1)\big)^2 (1 + 2 \alpha_m \Delta k) \nonumber \\
&\simeq \sum_i \big(\hat{r}_i(k_1)\big)^2 + 2 \alpha_m \Delta k \sum_i \big(\hat{r}_i(k_1)\big)^2.
\end{align}

Plugging this into Eq.~\ref{eq:MAPs_H0}, we find that
\begin{align}
\hat{H}_0(k_2) &= \frac{c \sum_i z_i \hat{r}_i(k_2)}{\sum_i \big(\hat{r}_i(k_2)\big)^2}
\simeq \frac{\sum_i z_i \hat{r}_i(k_1) (1 + \alpha_m \Delta k)}{\sum_i (\hat{r}_i(k_1))^2 (1 + 2 \alpha_m \Delta k)} \, c \nonumber \\
&\simeq \frac{\sum_i z_i \hat{r}_i(k_1)}{\sum_i (\hat{r}_i(k_1))^2} \, (1 - \alpha_m \Delta k) \, c \nonumber \\
&= \hat{H}_0(k_1) \, (1 - \alpha_m \Delta k).
\end{align}
Therefore, the first-order shift in $\hat{H}_0$ due to the prior change is
\begin{equation}\label{eq:H0_shift}
\Delta \hat{H}_0 \equiv \hat{H}_0(k_2) - \hat{H}_0(k_1) \simeq - \alpha_m \, \hat{H}_0(k_1) \, \Delta k.
\end{equation}
This matches Eq.~\ref{eq:deltaH0_app}, showing that whether the redshift or magnitude uncertainty dominates is not important for this result, but it cannot be expected to hold if they are comparable.

To express the bias in units of the $H_0$ uncertainty, we calculate the latter through linear error propagation. We can write Eq.~\ref{eq:MAPs_H0} as
\begin{equation}
\hat{H}_0 \;=\; c\,\frac{S_1}{S_2},
\end{equation}
where we have defined
\begin{equation}
S_1\equiv\sum_i z_i r_i,\qquad S_2\equiv\sum_i r_i^2.
\end{equation}
Differentiating with respect to $r_i$ and using $d r_i/d m_i = r_i/\alpha$, we obtain that
\begin{equation}
\frac{\partial \hat{H}_0}{\partial m_i}
= \frac{r_i}{\alpha}\;c\;\frac{z_i S_2 - 2 r_i S_1}{S_2^2}
= \frac{r_i}{\alpha S_2}\big(c z_i - 2 \hat{H}_0 r_i\big).
\end{equation}
Using that $cz_i \approx H_0 r_i$, this becomes
\begin{equation}
\frac{\partial \hat{H}_0}{\partial m_i}
\approx \frac{r}{\alpha S_2}\big(-\hat{H}_0 r\big)
= -\frac{\hat{H}_0}{\alpha}\,\frac{r^2}{S_2}.
\end{equation}
Now assuming that the objects are not at greatly different distances, we can take $S_2 \approx N r^2$ (in reality there will be an $\mathcal{O}(1)$ factor multiplying the right hand side) so that
\begin{equation}
\frac{\partial \hat{H}_0}{\partial m_i} \approx -\frac{\hat{H}_0}{\alpha}\,\frac{1}{N}.
\end{equation}
Hence the total contribution to the $H_0$ variance from the magnitude noise is
\begin{equation}
\sigma^2(H_0)_m \approx \sum_{i=1}^N \sigma_m^2
\Big(\frac{\partial \hat{H}_0}{\partial m_i}\Big)^2
\approx N\,\sigma_m^2\frac{\hat{H}_0^2}{\alpha^2}\frac{1}{N^2}
= \frac{\sigma_m^2}{\alpha^2}\frac{\hat{H}_0^2}{N}.
\end{equation}
The contribution from the redshifts is
\begin{align}
\sigma^2(H_0)_{z} &= \sum_{i=1}^N \left(\frac{\partial H_0}{\partial z_i}\right)^2 \sigma_z^2 \nonumber \\
&= c^2 \sigma_z^2 \frac{\sum_{i=1}^N {r}_i^2}{\left(\sum_{i=1}^N r_i^2\right)^2} = \frac{\hat{H}_0^2}{N} \frac{\sigma_z^2}{z_\text{eff}^2},
\end{align}
where we have defined
\begin{equation}
    z_\text{eff} \equiv \frac{\sum_i z_i r_i}{\sqrt{N \sum_i r_i^2}}.
\end{equation}

Combining the magnitude and redshift uncertainties, the total standard deviation is
\begin{equation}
\sigma(H_0) =\sqrt{\sigma(H_0)_m^2 + \sigma(H_0)_z^2} \simeq \frac{\hat{H}_0}{\sqrt{N}} \sqrt{ \alpha_m + \frac{\sigma_z^2}{z_\text{eff}^2}}.
\end{equation}
Combining with Eq.~\ref{eq:H0_shift}, this lets us calculate the first-order shift in $\hat{H}_0$ when changing the prior from $k_1$ to $k_2 \equiv k_1 + \Delta k$ as a relative bias of the posterior:
\begin{equation}\label{eq:mlimit_bias_rel}
\frac{\hat{H}_0(k_2)-\hat{H}_0(k_1)}{\sigma(H_0(k_1))} \simeq - \sqrt{N} \, \frac{\alpha_m \, \Delta k}{\sqrt{\alpha_m + \sigma_z^2 / z_\text{eff}^2}},
\end{equation}
implying the bias scales with $\sqrt{N}$.

As an example, if $\sigma_m = 0.1$, $\sigma_z = 0.001$, $N=2000$, and $r$ is uniformly distributed in volume between 5~Mpc and 100~Mpc, we have $\alpha_m = 0.00212$, $\langle z_\text{eff} \rangle = 0.0181$ for $H_0=70~\kmsecMpc$, $\sigma(H_0) = 0.113~\kmsecMpc$. This implies $\frac{\Delta \hat{H}_0}{\sigma(H_0)} \approx -1.32 \, \Delta k$, which corresponds to a 4.0$\sigma$ bias (high) in $H_0$ if the prior $\pi(r) \propto 1/r$ is used instead of the correct $r^2$ prior.

\end{appendix}

\bsp

\label{lastpage}
\end{document}